# Symmetries and pre-metric electromagnetism

**D.H. Delphenich**[*]

Physics Department, Bethany College, Lindsborg, KS 67456, USA

The equations of pre-metric electromagnetism are formulated as an exterior differential system on the bundle of exterior differential 2-forms over the spacetime manifold. The general form for the symmetry equations of the system is computed and then specialized to various possible forms for an electromagnetic constitutive law, namely, uniform linear, non-uniform linear, and uniform nonlinear. It is shown that in the uniform linear case, one has four possible ways of prolonging the symmetry Lie algebra, including prolongation to a Lie algebra of infinitesimal projective transformations of a real four-dimensional projective space. In the most general non-uniform linear case, the effect of non-uniformity on symmetry seems inconclusive in the absence of further specifics, and in the uniform nonlinear case, the overall difference from the uniform linear case amounts to a deformation of the electromagnetic constitutive tensor by the electromagnetic field strengths, which induces a corresponding deformation of the symmetry Lie algebra that was obtained in the linear uniform case.

**Contents**

---

[*] E-mail: delphenichd@bethanylb.edu

## 1  Introduction

It was recognized as early as 1910 by Bateman and Cunningham [**1, 2**] that the symmetries of Maxwell's equations for electromagnetism included not only the predictable Lorentz transformations, but also the conformal Lorentz transformations. More recently, in 1971 Harrison and Estabrook [**3**] derived this result in the context of the symmetries of exterior differential systems, a geometric approach to the representation of systems of partial differential equations on manifolds that had been largely established by Cartan [**4**] and Kähler [**5**] (cf., also Choquet-Bruhat [**6**]).  To be more precise, the symmetries of a system of partial differential equations refer to the transformations of the space of functions that the equations address that take solutions to other solutions.

    The primary purpose of this study is to formulate the pre-metric form of Maxwell's equations, in which the electromagnetic constitutive law of spacetime – essentially a scalar product on the bundle of 2-forms over the spacetime manifold, at least in the linear symmetric case – carries the fundamental geometric information instead of a Lorentzian structure on the tangent bundle, and then to investigate whether this expansion in the scope of the electrodynamical equations is associated with a resulting expansion in the scope of the symmetries of the equations.

    The topic of symmetries to the solutions of systems of differential equations – whether ordinary or partial – can be approached in various intersecting ways.  Some researchers (cf., e.g., [**7**]) prefer to regard a system of differential equations as simply a submanifold of an appropriate jet bundle that is defined by the null hypersurface of a set of functions on the bundle ([1]).  For instance, the two-dimensional Laplace equation:

$$u_{xx} + u_{yy} = 0 \qquad (1.1)$$

can be represented as the null hypersurface of the function:

$$f(x, y, u, p_x, p_y, q_{xx}, q_{xy}, q_{yy}) = q_{xx} + q_{yy} \qquad (1.2)$$

on the bundle $J^2(\mathbb{R}^2, \mathbb{R})$ of 2-jets of $C^2$ functions on $\mathbb{R}^2$.

    Others [**4, 5, 6**], preferring to follow the lead of É. Cartan, regard the system of differential equations as an exterior differential system − i.e., a finite set of exterior differential forms that vanish identically on any integral submanifold.  The Laplace equation can then be represented in various ways as a set of exterior differential forms. For instance, one might try to use the vanishing of the 0-form $\Delta u = \delta du$ to define the

---

[1] For other approaches to the problem of modeling systems of differential equations for the purpose of investigating their symmetries, cf. [**8-11**].

system, but the null hypersurface one would obtain is a submanifold of $\mathbb{R}^2$, not a set of harmonic 0-forms.

One could note that since:

$$du_x = u_{xx}\, dx + u_{xy}\, dy, \qquad du_y = u_{yx}\, dx + u_{yy}\, dy, \qquad (1.3)$$

then (1.1) can be expressed by the following exterior differential system on $J^1(\mathbb{R}^2, \mathbb{R})$:

$$0 = dp_x \wedge dy - dp_y \wedge dx, \qquad 0 = du - p_x\, dx - p_y\, dy. \qquad (1.4)$$

Here, we are looking at a sort of hybrid of the two approaches to the representation of partial differential equations, namely, an exterior differential system on a jet bundle ([2]). The appearance of the second equation is to restrict the solutions of the first equation to only those points of $J^1(\mathbb{R}^2, \mathbb{R})$ that are in the image of some 1-jet prolongation of a function in $C^2(\mathbb{R}^2)$, and one refers to the canonically-defined 1-form $du - p_x\, dx - p_y\, dy$ on $J^1(\mathbb{R}^2, \mathbb{R})$ as the *contact form* for $J^1(\mathbb{R}^2, \mathbb{R})$.

In some cases, it is not necessary to define one's exterior differential system on a jet bundle in order to represent a system of partial differential equations. For instance, in the case of the Laplace equation for a 1-form $\alpha = \alpha_x\, dx + \alpha_y\, dy$ on $\mathbb{R}^2$, the equation $\Delta\alpha = 0$ is equivalent to the pair of equations:

$$d\alpha = 0, \qquad \delta\alpha = 0, \qquad (1.5)$$

or:

$$d\alpha = 0, \qquad d{*}\alpha = 0, \qquad (1.6)$$

If one looks at a local trivialization $U \times \mathbb{R}^{2*}$ of the bundle $\Lambda^1(\mathbb{R}^2)$ whose coordinate functions take the form $(x, y, \alpha_x, \alpha_y)$ then if we take the simplest case, for which ${*}\alpha = -\alpha_y\, dx + \alpha_x\, dy$ (so $\alpha \wedge {*}\alpha = (\alpha_x)^2 + (\alpha_y)^2$), then since (1.6) takes the local form:

$$0 = \Theta^1 \equiv + d\alpha_x \wedge dx + d\alpha_y \wedge dy \qquad (1.7a)$$
$$0 = \Theta^2 \equiv - d\alpha_y \wedge dx + d\alpha_x \wedge dy, \qquad (1.7b)$$

one can also regard (1.6) as defining the exterior differential system (1.7a,b) on $\Lambda^1(\mathbb{R}^2)$.

One might suspect that since we have generalized $\alpha_x$ and $\alpha_y$ from functions on $\mathbb{R}^2$ to fiber coordinates on $\Lambda^1(\mathbb{R}^2)$ we would need to impose the conditions:

$$d\alpha_x = \alpha_{xx}\, dx + \alpha_{xy}\, dy, \qquad d\alpha_y = \alpha_{yx}\, dx + \alpha_{yy}\, dy, \qquad (1.8)$$

which are satisfied – indeed, they are *defined* – only when $\alpha_x$ and $\alpha_x$ are the components of a section of $\Lambda^1(\mathbb{R}^2)$ relative to the local coframe $dx$, $dy$. However, one finds that these

---

[2] Actually, a closer reading of Cartan from a modern perspective often shows that this is generally the case, since the supplementary variables that Cartan introduced in order to prolong a non-integrable exterior differential system into an integrable one were usually just higher derivatives of the component functions; hence, one was simply prolonging the exterior differential system to a higher order jet bundle.

conditions are simply a consequence of assuming that $\alpha$ is a *solution* to the system (1.7), namely:

$$\alpha^* \Theta^i = 0, \qquad i = 1, 2, \tag{1.9}$$

since the process of pulling back the 2-forms $\Theta^i$ will replace $d\alpha_x$ and $d\alpha_y$ with the respective expressions in (1.8). Hence, we can treat the system (1.7a,b) as an adequate representation of the Laplace equation for 1-forms by an exterior differential system on $\Lambda^1(\mathbb{R}^2)$.

In the next section we briefly summarize the relevant notions from the theory of exterior differential systems, largely for the sake of completeness of the presentation. Then we discuss the symmetries of exterior differential systems and apply the formalism of the first two subsections to the case at hand of the total space of the bundle of exterior differential 2-forms on the spacetime manifold. We then present the equations of pre-metric electromagnetism as an exterior differential system on $\Lambda^2(M)$, the bundle of 2-forms on the spacetime manifold $M$, and find the general form for its infinitesimal symmetry generators. Finally, we examine some special cases that follow from possible forms of the electromagnetic constitutive law.

## 2 Exterior differential systems [4-6]

Since we shall eventually represent the pre-metric form of the Maxwell equations as an exterior differential system on the total space to the vector bundle $\Lambda^2(M) \to M$ of exterior differential 2-forms on the spacetime manifold $M$, we briefly recall the concepts that are most relevant to that objective.

### 2.1 Basic concepts

A *differential system* ([3]) on an $n$-dimensional differentiable manifold $M$ is a sub-bundle $\mathcal{D}(M)$ of the tangent bundle $T(M)$ whose fibers have constant rank (i.e., dimension) – say $k$. Hence, at every point $x \in M$ one has a $k$-dimensional linear subspace $\mathcal{D}_x(M)$ of $T_x(M)$. One can also regard a rank-$k$ differential system on $M$ as a section of the tangent Grassmannian bundle $GT^{k,n}(M)$, whose fiber at each $x \in M$ is the Grassmanian manifold of all $k$-dimensional linear subspaces of $T_x(M)$. For instance, a rank-one differential system on $M$ is a global line field, which can be regarded as a section of $PT(M)$, the projectivized tangent bundle.

The immediate issue that one encounters once one has defined such a differential system is its integrability. The rank-$k$ differential system $\mathcal{D}(M)$ is *completely integrable* ($^4$) iff there is a foliation of $M$ by $k$-dimensional *integral submanifolds*; i.e., submanifolds of $M$ whose $k$-dimensional tangent spaces all agree with the fibers of $\mathcal{D}(M)$.

---

[3] The term *distribution* is also used by some authors, but it will not be used here since it can also be used to describe linear functionals on a vector space of sections of a vector bundle, such as the bundle of exterior differential forms.

If we denote the vector space of all sections of the vector bundle $\mathcal{D}(M)$ by $\mathfrak{X}_\mathcal{D}(M)$, i.e., the space of all vector fields on $M$ that take their values in the fibers of $\mathcal{D}(M)$, then the first form of Frobenius's theorem says that $\mathcal{D}(M)$ is completely integrable iff $\mathfrak{X}_\mathcal{D}(M)$ is a sub-algebra of $\mathfrak{X}(M)$ under the Lie bracket of vector fields. One then says that $\mathcal{D}(M)$ is *involutory*.

One way of singling out a particular sub-bundle $\mathcal{D}(M)$ of $T(M)$ is to make $\mathcal{D}(M)$ the solution of a finite set of algebraic equations that take the form of the vanishing of a finite set of exterior differential forms $\mathcal{E} = \{\Theta^i \in \Lambda^{k_i}(M), i = 1, \ldots, m\}$ on $M$:

$$0 = \Theta^i(\mathbf{v}) \quad \text{for all } i \tag{2.1}$$

when and only when they are evaluated on any $k_i$-vector $\mathbf{a} \in \Lambda_{k_i}(\mathcal{D})$ as *integral elements* of the system.

When the differential system $\mathcal{D}(M)$ is obtained in this way, one arrives at the second form of Frobenius's theorem that $\mathcal{D}(M)$ is completely integrable iff for each $i$ one has:

$$d\Theta^i = \eta^i_j \wedge \Theta^j \tag{2.2}$$

for some set of 1-forms $\eta^i_j$. Clearly, it is sufficient, but not necessary, that $\Theta^i$ be closed.

Finally, since (2.1) is not the only system of $m$ differential forms that will give the same integral elements as it does, one can characterize an exterior differential system as the *ideal I* in the exterior algebra $\Lambda^*(M)$ that is *generated* by the set $\mathcal{E} = \{\Theta^i, i = 1, \ldots, m\}$. This means that $I$ consists of all differential forms on $M$ that take the form of finite linear combinations of terms that are formed from exterior products of the elements of $\mathcal{E}$ and any other differential forms on $M$. For instance, one might represent a typical element (using the summation convention) as:

$$\alpha_{ij\ldots k} \, \eta^i \wedge \Theta^j \wedge \ldots \wedge \Theta^k. \tag{2.3}$$

Hence, one can give a third form of Frobenius's theorem by regarding (2.2) as the statement that the ideal $I$ is closed under exterior differentiation:

$$dI \subset I. \tag{2.4}$$

Something that becomes apparent in this form is that if the ideal generated by $\mathcal{E}$ is not closed under $d$ then all one has to do to make it closed is to add more generators, in the form of all of the $d\Theta^i$. We then denote the extended set of generators by $\overline{\mathcal{E}} = \{\Theta^i, d\Theta^i, i = 1, \ldots, m\}$ and the refer to the ideal $\overline{I}$ that is generated by $\overline{\mathcal{E}}$ as the *closure* of $I$. One also notes that the exterior differential system (2.1) must be augmented to:

---

[4] The reason that one includes the adjective "completely" is because it is conceivable that one might have integral submanifolds of dimension less than $k$, but not of dimension $k$.

$$\begin{cases} 0 = \Theta^i \\ 0 = d\Theta^i \end{cases} \text{ for all } i. \tag{2.5}$$

## 2.2 Exterior differential systems on $\Lambda^2(M)$

We now replace the manifold $M$ on which we define the differential system $\mathcal{D}(M)$ with the manifold $\Lambda^2(M)$; i.e., the total space of the vector bundle of exterior differential 2-forms on $M$. This is because a solution to our field equations will take the form of a section $F: M \to \Lambda^2(M)$ of this vector bundle, so $F$ can also be regarded as a special type of four-dimensional submanifold of $\Lambda^2(M)$ that is transverse to the fibers and which projects back to itself identically.

We wish to represent $F$ as an integral manifold to an exterior differential system $I$ on $\Lambda^2(M)$ that is generated by a set $\mathcal{E} = \{\Theta^i, i = 1, \ldots, m\}$ of elements of $\Lambda^*(\Lambda^2(M))$. Hence, that system must then have four-dimensional integral elements that are transverse to the vertical subspaces at each point of $\Lambda^2(M)$.

Here, we need to confront the fact that not all four-dimensional submanifolds of the form $\phi: M \to \Lambda^2(M)$ represent *sections*, but only the ones for which $\pi \cdot \phi = I$, where $\pi$ is the bundle projection. This differentiates to the global condition on the differential maps:

$$D\pi \cdot D\phi = I. \tag{2.6}$$

If a local trivialization of $\Lambda^2(M)$ over a coordinate chart $\{U, x^\mu\}$ looks like $(x^\nu, F_{\mu\nu})$, so $\pi(x^\nu, F_{\mu\nu}) = x^\nu = [I \mid 0] [x^\nu, F_{\mu\nu}]^T$ and $\phi(x) = (\phi^\mu(x^\lambda), \phi_{\mu\nu}(x^\lambda))$, then:

$$D\pi|_{(x, F)} = [I \mid 0], \qquad D\phi = [\partial_\lambda \phi^\mu \mid \partial_\lambda \phi_{\mu\nu}], \tag{2.7}$$

which makes the condition (2.6) take the local form:

$$\frac{\partial \phi^\mu}{\partial x^\lambda} = \delta^\mu_\lambda, \tag{2.8}$$

which integrates to the condition:

$$\phi^\mu = x^\mu + \varepsilon^\mu. \tag{2.9}$$

Hence, $\phi: M \to \Lambda^2(M)$ is a section iff for any local trivialization of $\Lambda^2(M)$ over a coordinate chart on $M$, $\phi$ takes the form $\phi(x) = (x^\lambda, \phi_{\mu\nu}(x^\lambda))$.

Since a local trivialization of $\Lambda^2(M)$ looks like $(x^\mu, F_{\mu\nu})$, an arbitrary vector field on $\Lambda^2(M)$ will take the local form:

$$X = X^\mu \frac{\partial}{\partial x^\mu} + X_{\mu\nu} \frac{\partial}{\partial F_{\mu\nu}}, \tag{2.10}$$

in which $X^\mu, X_{\mu\nu} \in C^\infty(\Lambda^2(M))$ – i.e., $X^\mu = X^\mu(x^\mu, F_{\mu\nu})$ and $X_{\mu\nu} = X_{\mu\nu}(x^\mu, F_{\mu\nu})$.

The exterior algebra $\Lambda^*(\Lambda^2(M))$ is locally generated by the set $\{dx^\mu, dF_{\mu\nu}\}$, which has $n + \frac{1}{2}n(n-1) = \frac{1}{2}n(n+1)$ elements in it, so that will be the maximum $k$ for which there are non-zero $k$-forms on $\Lambda^*(\Lambda^2(M)$; in particular, when $n = 4$, this will be equal to 10.

### 2.3 Canonical forms on $\Lambda^2(M)$

There is a canonical 1-form $\theta$ on $\pi\colon T^*(M) \to M$ that is defined by the condition:

$$\theta_p(\mathbf{v}) = p_x(\pi_*\mathbf{v}) \tag{2.11}$$

whenever $\mathbf{v} \in T_p(T^*_x(M))$.

One can analogously define a canonical 2-form $\Phi$ on $\Lambda^2(M)$ ([5]). If $x \in M$, $F \in \Lambda^2_x(M)$, $\mathbf{v}, \mathbf{w} \in T_F(\Lambda^2_x(M))$ then one simply sets:

$$\Phi(\mathbf{v}, \mathbf{w}) = F(\pi_*\mathbf{v}, \pi_*\mathbf{w}), \tag{2.12}$$

in which, by abuse of notation, we are using the same symbol for the bundle projection as before.

For a local trivialization $\Lambda^2(U) = U \times \Lambda^2(\mathbb{R}^4)$ of $\Lambda^2(M)$ over a coordinate chart $(U, x^\mu)$ the canonical 2-form $\Phi$ takes the local form:

$$\Phi = \tfrac{1}{2} F_{\mu\nu} dx^\mu \wedge dx^\nu. \tag{2.13}$$

We hasten to emphasize that this is not the same thing as the local expression for a 2-form on $U$ since the components of $\Phi$ in (2.13) are not functions of $U$, but, in fact, coordinate functions on $\Lambda^2(U)$. However, for a section $F\colon M \to \Lambda^2(M)$, the canonical 2-form $\Phi$ pulls down to the 2-form $F$ itself, and if $F$ is a local section over $U \subset M$ then the component functions $F_{\mu\nu}$ become functions on $U$.

On first glance, it is conceivable that the annihilating subspaces of $\Phi$ in $T(\Lambda^2(M))$ do not have constant dimension. The basic issue is how the rank of the 2-form $\Phi$ at $F \in \Lambda^2_x(M)$ relates to the rank of $F$.

There are various equivalent definitions of rank, but for our immediate purposes, we shall think of the rank of $F$ or $\Phi$ as the codimension of the associated subspaces of $F$ or $\Phi$ in $T(M)$ or $T(\Lambda^2(M))$, respectively. For instance, the associated subspace of $F$ at $x \in M$ will consist of all tangent vectors $\mathbf{v} \in T_x(M)$ such that $i_\mathbf{v} F = 0$, which called the *associated equation* for $F$ at $x$. Note that although this subspace will be *contained in* the annihilating subspace for $F$ at $x$, it will not necessarily *agree* with it, unless $F$ is decomposable. As a counter-example, consider an $F$ that is not decomposable and takes the form $F = \alpha \wedge \beta + \gamma \wedge \delta$, with all covectors linearly independent. One can easily find vectors $\mathbf{v}, \mathbf{w}$ such that $0 = F(\mathbf{v}, \mathbf{w}) = \alpha \wedge \beta(\mathbf{v}, \mathbf{w}) + \gamma \wedge \delta(\mathbf{v}, \mathbf{w})$, even though the two terms are not individually zero. Indeed, since $\alpha, \beta, \gamma, \delta$ are linearly independent they define a coframe for $T_x^*(M)$ at each $x \in M$. Hence, they also define a reciprocal frame $\mathbf{a}, \mathbf{b}, \mathbf{c}, \mathbf{d}$ in each $T_x(M)$. Now:

$$F(\mathbf{a} + \mathbf{c}, \mathbf{b} - \mathbf{d}) = \alpha(\mathbf{a})\beta(\mathbf{b}) - \gamma(\mathbf{c})\delta(\mathbf{d}) = 1 - 1 = 0. \qquad (2.14)$$

---

[5] Indeed, from the definition (2.12), it is clear the there is, more generally, a canonical $k$-form on any bundle $\Lambda^k(M)$ of $k$-forms. The canonical 1-form on $T^*(M)$ is a special case of this if one regards $T^*(M)$ as the same thing as $\Lambda^1(M)$.

Therefore, the 2-vector $(\mathbf{a} + \mathbf{c}) \wedge (\mathbf{b} - \mathbf{d})$ annihilates $F$ even though:

$$i_{(\mathbf{a}+\mathbf{c})} F = \beta + \delta \neq 0, \qquad i_{(\mathbf{b}-\mathbf{d})} F = \alpha - \gamma \neq 0; \qquad (2.15)$$

Thus, neither $\mathbf{a} + \mathbf{c}$ nor $\mathbf{b} - \mathbf{d}$ is an element of the associated space to $F$ at $x$.

The latter counter-example also shows that the dimension of the annihilating subspace of $F$ at $x$ is always at least two, since the vectors $\mathbf{a} + \mathbf{c}$ and $\mathbf{b} - \mathbf{d}$ are linearly independent, and will therefore span a 2-plane in $T_x(M)$, all of whose 2-vectors will annihilate $F$. Of course, when $F = 0$, the dimension of the annihilating subspace is four, but the question arises whether the dimension is ever greater than two for non-zero $F$; in particular, for $F$ of rank four. However, if one regards an annihilating subspace for $F$ as also a "maximal isotropic" subspace for $F$ then one finds that since the dimension of such subspaces must be one-half the dimension of the space when $F$ is non-zero, the dimension two is also maximal. Hence, the annihilating subspaces of $F$ will be two-dimensional whenever $F$ is non-zero.

This implies that the annihilating subspaces of $\Phi$ will be eight-dimensional, since they will consist of the direct sum of a copy of the maximal isotropic subspaces of $F$ in the $T_F(\Lambda^2(M))$ and the vertical subspaces of $T_F(\Lambda^2(M))$. However, it is important to notice that the association of a maximal isotropic subspace of $T_x(M)$ for $F_x$ with a corresponding "non-vertical" subspace of $T_F(\Lambda^2(M))$ will not be unique, and will generally depend upon a choice of connection on $\Lambda^2(M)$ that would define a horizontal complement to the vertical sub-bundle of $T(\Lambda^2(M))$.

One sees that $\Phi$ gives us a canonical 3-form:

$$d\Phi = \tfrac{1}{2} dF_{\mu\nu} \wedge dx^\mu \wedge dx^\nu. \qquad (2.16)$$

As we shall see, the canonical 3-form $d\Phi$ is one of the two 3-forms on $\Lambda^2(M)$ that collectively define the generators of the exterior differential system that represents the pre-metric form of Maxwell's equations. One sees that the solutions of the exterior differential system that $d\Phi$ defines, which will be submanifolds of $\Lambda^2(M)$ on which the restriction of $d\Phi$ vanishes, will include all closed 2-forms on $M$. However, since $d\Phi$ is a 3-form on a ten-dimensional manifold, the annihilating subspaces of $d\Phi$ will be seven-dimensional, which implies that there will be "more" integral submanifolds for the exterior differential system $d\Phi = 0$ than closed 2-forms, since the image of $M$ by a section of $\Lambda^2(M)$ will only be a four-dimensional submanifold.

It is also important to consider the rank singularities of $F$, $\Phi$, and $d\Phi$ since they will affect the infinitesimal symmetries of those forms by way of the Lie derivative.

Since $F$ and $\Phi$ are both 2-forms they will have even rank in either case. Hence, the possible ranks for $F$ are 0, 2, and 4, whereas $\Phi$ can have rank 0, 2, …, 10.

Let us start with $F$: Rank 0 is the trivial case of $F = 0$, which means that the associated subspace $\text{Ann}_x(F)$ in any fiber $T_x(M)$ is the whole fiber. Rank two means that there will be non-zero vectors **v** in $T(M)$ that satisfy the associated equation of $F$:

$$i_\mathbf{v} F = 0 , \qquad (2.17)$$

or, in local components:

$$v^\mu F_{\mu\nu} = 0 , \qquad (2.18)$$

but, unlike the rank 0 case, not all tangent vectors will satisfy this system of algebraic equations. Hence, the dimension of the associated subspace, as well as the annihilating subspace, will be two. Finally, in the case of rank 4, the only solution to (2.17) will be the zero vector in each fiber.

The singular $F$ have great significance in electromagnetism, since they include the wavelike solutions of the Maxwell equations as special cases. When one expresses a 2-form $F$ in $E$-$B$ form by making a choice of timelike unit vector field **t**, so $E = i_\mathbf{t} F$ and $B = i_\mathbf{t}*F$, so $F = \theta \wedge E - *(\theta \wedge B)$, with $\theta = i_\mathbf{t} g$, then one sees that $F \wedge F = 2(E \cdot B)\,\mathcal{V}$, and the rank-two 2-forms are characterized by the simple property that $E$ is spatially orthogonal to $B$. Hence, they will span a 2-plane in every cotangent space, as well as every tangent space, by way of their annihilating vectors. When $E$ and $B$ define an electromagnetic wave, one refers to this 2-plane as the *plane of polarization* of the wave.

As for the rank of $\Phi$, one notes from either (2.17) or (2.18) that any tangent vector to $\Lambda^2(M)$ that is vertical – i.e., tangent to the fiber – will be annihilated by $\Phi$. Hence, we know that the dimension of the associated subspace to $\Phi$ at a given point $F \in \Lambda^2_x(M)$ will be at least 6. One then sees that whether the total dimension of the associated subspace at $F$ is 6, 8, or 10 depends upon whether the rank of $F$ is 4, 2, or 0, respectively.

We should point out that the rank of $\Phi$ is, in a sense, "generically" six, since in each fiber of $\Lambda^2(M)$ there is only one 2-form that has rank 0 and the space of 2-forms with rank 2 is a hypersurface – the *Klein quadric* – that separates the two remaining open submanifolds of the fiber that consist of 2-forms of rank four. Hence, "most" $F$ are non-singular points for $\Phi$, at which the associated subspace for $\Phi$ is simply the vertical subspace of $\Lambda^2(M)$. These $F$ collectively define a singular foliation of $\Lambda^2(M)$ whose leaves consist of the fibers of $\Lambda^2(M)$ minus the Klein quadric.

The associated spaces for $d\Phi$ will consist of vectors $X$ on $\Lambda^2(M)$ that satisfy:

$$i_X d\Phi = 0 , \qquad (2.19)$$

which takes the local form:

$$0 = X_{\mu\nu}\, dx^\mu \wedge dx^\nu - 2X^\mu\, dF_{\mu\nu} \wedge dx^\nu . \qquad (2.20)$$

Since the $dx^\mu \wedge dx^\nu$ and the $dF_{\mu\nu} \wedge dx^\nu$ are linearly independent the only solution to (2.20) is the zero vector field on $\Lambda^2(M)$; hence, the rank of $d\Phi$ is everywhere maximal, i.e., ten.

## 3 Symmetries of exterior differential systems

The transformations that act on a manifold that will be of greatest interest to us in the present study will be the ones that take solutions of the exterior differential system that we will be dealing with – viz., the pre-metric form of the Maxwell equations – to other solutions. Although if one regards a solution as a special type of section of the vector bundle $\Lambda^2(M) \to M$ in that particular case this suggests that we shall be concerned with groups that act on this infinite-dimensional vector space of sections, nevertheless, we shall not follow this course of action, but will concentrate instead on groups that act on the finite-dimensional total space of $\Lambda^2(M)$.

We shall now try to make all of this more precise.

### 3.1 Basic concepts

If $\mathcal{D}(M)$ is a completely integrable differential system on $M$ then a *symmetry* of $\mathcal{D}(M)$ is a diffeomorphism $f: M \to M$ that takes any leaf $L$ of the foliation of $M$ by integral submanifolds to another leaf $\phi(L)$ of that foliation; i.e., it takes solutions of $\mathcal{D}(M)$ to other solutions. Note that there are basically two ways that this can happen: either $\phi(L) = L$ or $\phi(L)$ is some other (diffeomorphic) leaf of the foliation.

If one has a differentiable one-parameter family $\phi_s$, $s \in \mathbb{R}$ of such diffeomorphisms and one differentiates with respect to the parameter $s$ then one obtains a vector field $X$ on the domain of $\phi_s$ that represents an infinitesimal generator of the one-parameter family of symmetries of $\mathcal{D}(M)$; hence, one refers to $X$ as an *infinitesimal symmetry* of the differential system $\mathcal{D}(M)$. If the one-parameter family of symmetries that $X$ generates consists of leaf-preserving diffeomorphisms then we say that $X$ is *vertical to the foliation*, or *tangent to the leaves* ([6]); otherwise, we say it is *transverse* to the foliation. This implies that $X$ is vertical to the foliation iff $X \in \mathfrak{X}_{\mathcal{D}}(M)$. For example, for a one-dimensional differential system these transformations amount to reparameterizations of the integral curve, at least locally.

If $Y \in \mathfrak{X}_{\mathcal{D}}(M)$ is a vector field on $M$ that is tangent to the foliation defined by $\mathcal{D}(M)$ then $X$ is an infinitesimal symmetry of $\mathcal{D}(M)$ iff:

$$[X, Y] \in \mathfrak{X}_{\mathcal{D}}(M). \tag{3.1}$$

Note that this does not have to imply that $X \in \mathfrak{X}_{\mathcal{D}}(M)$; e.g., $X$ can be transverse to the leaves.

Now, suppose that the fibers of $\mathcal{D}(M)$ are the integral elements of some exterior differential system on $M$, which we regard as a closed ideal $I$ in $\Lambda^*(M)$ that is generated by the elements $\{\theta^1, \ldots, \theta^m\}$. In order for a diffeomorphism $\phi: M \to M$ to take leaves to leaves, its differential map must also take integral elements to integral elements:

$$\phi_*(\mathcal{D}_x(M)) = \mathcal{D}_{\phi(x)}(M) \tag{3.2}$$

---

[6] Since a fibration is a special type of foliation this is not particularly an abuse of terminology.

for all $x \in M$. Hence, it must also pull back the restriction $I_{\phi(x)}$ of the ideal $I$ to the point $\phi(x)$ to its restriction $I_x$ at $x$:

$$I_x = \phi^*(I_{\phi(x)}) \tag{3.3}$$

for all $x \in M$. In order for this to be true it is necessary and sufficient that it be true for all the generators $\mathcal{E}$ of $I$. Hence, one can also say that:

$$\phi^*(\mathcal{E}) = \mathcal{E}. \tag{3.4}$$

In the case of an infinitesimal symmetry of $\mathcal{D}(M)$, one then has the following condition, which is equivalent to (3.1):

$$L_X \mathcal{E} \in I, \tag{3.5}$$

i.e., for each generator $\theta^i \in \mathcal{E}$ one must have:

$$L_X \theta^i \in I. \tag{3.6}$$

We shall also use the notation $L_X \theta^i \equiv 0 \pmod{I}$ when we wish to convey the same meaning as (3.6).

### 3.2 Some transformations that act on $\Lambda^2(\mathbb{R}^4)$

Although the subject of transformations that act on $M$ and $\Lambda^2(M)$ globally has considerable merit in the context of cosmological models with gravitational fields, nevertheless, the symmetries of electromagnetic fields often tend to have a local character that makes it appropriate to consider only local actions. Hence, we shall regard $\Lambda^2(M)$ as essentially represented by $U \times \Lambda^2(\mathbb{R}^4)$, where $U$ is an open subset of $M$ that is diffeomorphic to $\mathbb{R}^4$ by way of the coordinate map $x^\mu$, so we shall really be looking at group actions on $\mathbb{R}^4 \times \Lambda^2(\mathbb{R}^4)$, with coordinate functions $x^\mu$ on $\mathbb{R}^4$ and $F_{\mu\nu}$ on $\Lambda^2(\mathbb{R}^4)$. We shall first examine transformations of $\mathbb{R}^4$ and then transformations of $\Lambda^2(\mathbb{R}^4)$.

Since both $\mathbb{R}^4$ and $\Lambda^2(\mathbb{R}^4)$ are vector spaces, we point out that one will always have the action of $\mathbb{R}$ on either space by scalar multiplication and the action of the vector space on itself by translation. As a consequence, any system of linear differential equations will admit the symmetries that arise from the fact that a linear combination of a given solution and any other solution must still be a solution.

### 3.2.1 Transformations of $\mathbb{R}^4$

There are two basic types of group actions on any vector space: linear and nonlinear. A linear action $G \times \mathbb{R}^4 \to \mathbb{R}^4$ of a group $G$ on $\mathbb{R}^4$ is equivalent to a representation of the group $G$ in $GL(4; \mathbb{R})$; i.e., a homomorphism $D: G \to GL(4; \mathbb{R})$.

By differentiation at the identity if $G$, we also see that a homomorphism of $G$ into $GL(4; \mathbb{R})$ also gives us a homomorphism of the Lie algebras $\mathcal{D}: \mathfrak{g} \to \mathfrak{gl}(4; \mathbb{R})$. Now, the action of $\mathfrak{gl}(4; \mathbb{R})$ is really on tangent vectors to $\mathbb{R}^4$, not the vectors themselves − although there is a canonical linear isomorphism of $\mathbb{R}^4$ with any of its tangent spaces − and this induces a homomorphism of $\mathfrak{g}$ into $\mathfrak{X}(\mathbb{R}^4)$, the Lie algebra of vector fields on $\mathbb{R}^4$. This homomorphism simply takes $a \in \mathfrak{g}$ to the corresponding fundamental vector field $\tilde{a}$ of the action of the one-parameter subgroup $\exp(as) \in G$.

The linear actions that we shall consider are all described by representations of the subgroups of $GL(4; \mathbb{R})$ in which the homomorphism that defines the representation is simply the subgroup inclusion. Hence, we discuss the action of $GL(4; \mathbb{R})$ on $\mathbb{R}^4$ and then simply specialize the result to some of its relevant subgroups.

The linear action of $GL(4; \mathbb{R})$ on $\mathbb{R}^4$ that we shall consider is the simplest non-trivial one, viz., the defining representation, which simply means that if the elements of $GL(4; \mathbb{R})$ are described by invertible real 4×4 matrices and the elements of $\mathbb{R}^4$ are described by real 4×1 column matrices then the action of $GL(4; \mathbb{R})$ on $\mathbb{R}^4$ is by matrix multiplication:

$$GL(4; \mathbb{R}) \times \mathbb{R}^4 \to \mathbb{R}^4, \qquad (A^\mu_\nu, x^\nu) \mapsto A^\mu_\nu x^\nu. \qquad (3.7)$$

If we now consider a one-parameter subgroup of $GL(4; \mathbb{R})$ of the form:

$$A^\mu_\nu(s) = \exp(a^\mu_\nu s) \qquad (3.8)$$

with $a^\mu_\nu \in \mathfrak{gl}(4; \mathbb{R})$ then we can obtain the components of the fundamental vector field on $\mathbb{R}^4$ that has $a^\mu_\nu$ for its infinitesimal generator by differentiation:

$$X^\mu(x^\lambda) = \left.\frac{d}{ds}\right|_{s=0} \exp(a^\mu_\nu s) x^\nu = a^\mu_\nu x^\nu ; \qquad (3.9)$$

hence:

$$X(x^\lambda) = a^\mu_\nu x^\nu \frac{\partial}{\partial x^\mu}. \qquad (3.10)$$

We shall now restrict this to various subgroups of $GL(4; \mathbb{R})$.

$SL(4; \mathbb{R})$: An element $A^\mu_\nu \in SL(4; \mathbb{R})$ differs from the general element of $GL(4; \mathbb{R})$ only by the fact that $\det(A^\mu_\nu) = 1$. This means that the action of $A^\mu_\nu$ on $\mathbb{R}^4$ will preserve any chosen unit-volume element $\mathcal{V} \in \Lambda^4(\mathbb{R}^4)$ on $\mathbb{R}^4$. An element $a^\mu_\nu$ of the Lie algebra $\mathfrak{sl}(4;\mathbb{R})$, which defines an infinitesimal generator of a volume-preserving transformation, will then differ from the general element of $\mathfrak{gl}(4;\mathbb{R})$ by the fact that $\text{Tr}(a^\mu_\nu) = a^\mu_\mu = 0$. As a result the fundamental vector field that it generates on $\mathbb{R}^4$ will have the property ([7]):

$$\delta X = \#^{-1}d\#X = \frac{\partial X^\mu}{\partial x^\mu} = a^\mu_\nu \frac{\partial x^\nu}{\partial x^\mu} = a^\mu_\nu \delta^\nu_\mu = a^\mu_\mu = 0. \qquad (3.11)$$

---

[7] Note, in particular, that it is possible to define the divergence of a vector field using only a unit-volume element; it is not necessary to introduce a metric.

Hence, the infinitesimal generator of the action of a one-parameter subgroup of *SL*(4; ℝ) on ℝ$^4$ will be a *conserved current* on ℝ$^4$, i.e., a divergenceless vector field on ℝ$^4$.

The elements of *SL*(4; ℝ) differ from the general elements of *GL*(4; ℝ) only by a non-zero scalar multiplication, which really just amounts to the factorization of $A_\nu^\mu \in GL(4; ℝ)$ into:

$$A_\nu^\mu = \det(A_\nu^\mu)^{1/4} \, [\det(A_\nu^\mu)^{-1/4} A_\nu^\mu] = \alpha \, \tilde{A}_\nu^\mu, \tag{3.12}$$

so $\alpha \in ℝ^*$ and $\tilde{A}_\nu^\mu \in SL(4; ℝ)$. Hence, we should also describe the action of $\alpha$ on ℝ$^4$, which is referred to as a *dilatation* or *homothety*. One simply represents the scalar $\alpha$ as the matrix $\exp(\lambda)\delta_\nu^\mu \in GL(4; ℝ)$ and specializes (3.10) to:

$$X = \lambda \, x^\mu \, \frac{\partial}{\partial x^\mu}. \tag{3.13}$$

*O*(3, 1): When one gives ℝ$^4$ an orthogonal structure by way of the Minkowski scalar product, whose components with respect to the canonical basis are $\eta_{\mu\nu} = \mathrm{diag}(+1, -1, -1, -1)$, one can identify the subgroup *O*(3, 1) of *GL*(4; ℝ) that consists of all invertible real 4×4 matrices that preserve $\eta$; i.e., all *A* such that:

$$A^T \eta A = \eta. \tag{3.14}$$

This subgroup is referred to as the *Lorentz group*.

The defining condition (3.14) for elements of *O*(3, 1) differentiates to the defining condition for the elements *a* of the Lie algebra $\mathfrak{so}(3, 1)$, namely:

$$a^T \eta + \eta a = 0, \tag{3.15}$$

or:

$$a_{\nu\mu} = -a_{\mu\nu}, \tag{3.16}$$

where have defined $a_{\mu\nu} = \eta_{\mu\lambda} \, a_\nu^\lambda$. Hence, the infinitesimal generator of any Lorentz transformation can be associated with a unique anti-symmetric matrix.

The Lie algebra $\mathfrak{so}(3,1)$ decomposes as a vector space (but not as a Lie algebra) into a three-dimensional subspace $\mathfrak{so}(3)$ of infinitesimal Euclidian rotations, which *is* a subalgebra of $\mathfrak{so}(3,1)$, and a three-dimensional vector space $\mathfrak{b}$ of infinitesimal Lorentz boosts, which is not.

A particular elementary example of such a transformation is a Euclidian rotation in the *xy*-plane. Its infinitesimal generator can be represented in the form:

$$a_j^i = J_j^i = \begin{bmatrix} 0 & -1 \\ 1 & 0 \end{bmatrix}. \tag{3.17}$$

The corresponding fundamental vector field on ℝ$^4$ is then:

$$X = y\frac{\partial}{\partial x} - x\frac{\partial}{\partial y}. \tag{3.18}$$

By comparison to (3.17), an infinitesimal boost in the *xt*-plane, which takes the form:

$$a^i_j = \Sigma^i_j = \begin{bmatrix} 0 & 1 \\ 1 & 0 \end{bmatrix}, \tag{3.19}$$

corresponds to the fundamental vector field on $\mathbb{R}^4$:

$$X = x\frac{\partial}{\partial t} + t\frac{\partial}{\partial x}. \tag{3.20}$$

*CO*(3, 1): The *conformal Lorentz group CO*(3, 1) is the subgroup of Diff($\mathbb{R}^4$) that consists of diffeomorphisms *A* such that there is some $\Omega \in \mathbb{R}^*$, which will depend on *A*, such that:

$$A^T \eta A = \Omega(A)\,\eta. \tag{3.21}$$

Clearly, the Poincaré group is the subgroup of this group for which $\Omega(A)$ always equals 1. The only types of elements that a different choice of $\Omega(A)$ would introduce are homotheties and inversions, which we will discuss shortly. However, we must emphasize at this point that the only subgroup of the affine group *A*(4; $\mathbb{R}$) that (3.21) defines is the direct product of the homothety group and the Poincaré group, a group that one calls the *affine conformal Poincaré group;* without the translations, one defines the *linear conformal Lorentz group*.

For the linear elements in *CO*(3, 1), which can be represented by constant matrices in *GL*(4; $\mathbb{R}$), (3.21) differentiates at the identity to the defining condition for elements of the linear conformal Lorentz Lie algebra $\mathfrak{co}(3, 1)$, that for every $a \in \mathfrak{co}(3, 1)$ there should be some $\lambda(a) \in \mathbb{R}$ such that:

$$a^T \eta + \eta a = \lambda(a)\eta. \tag{3.22}$$

If one rewrites (3.22) as:

$$a_{\nu\mu} + a_{\mu\nu} = \lambda(a)\eta_{\mu\nu} \tag{3.23}$$

then since the left-hand side is twice the symmetric part of $a_{\mu\nu}$, we see that (3.22) tells us that the anti-symmetric part of $a_{\mu\nu}$ is arbitrary, but the symmetric part is $\lambda(a)\eta_{\mu\nu}$. Hence, upon raising an index in (3.23), we see that the arbitrary element of $\mathfrak{co}(3, 1)$ takes the form:

$$a^\mu_{\ \nu} = \omega^\mu_{\ \nu} + \alpha\,\delta^\mu_{\ \nu}, \tag{3.24}$$

with $\omega^\mu_{\ \nu} \in \mathfrak{so}(3, 1)$ and $\alpha \in \mathbb{R}$. Hence, as a Lie algebra $\mathfrak{co}(3, 1) = \mathfrak{so}(3, 1) \oplus \mathbb{R}$.

This does not, however, define the entire conformal Lorentz Lie algebra, since we must also account for the transformations that act nonlinearly on $\mathbb{R}^4$, as well, namely, the translations and inversions through the light cone. We shall eventually assemble all of these pieces in the context of the prolongations of Lie algebras, but first we discuss the translations and inversions by themselves.

In general, a nonlinear action of a group on $\mathbb{R}^4$ will be represented by a homomorphism of a group $G$ into Diff($\mathbb{R}^4$). A simple way that this can come about is when $G$ is defined as a subgroup of Diff($\mathbb{R}^4$) to begin with, as we did with the conformal Lorentz group. Although nonlinear actions in their most general form can get very abstruse and esoteric, we shall confine ourselves to two particular nonlinear actions whose nonlinearity comes from the projection of a linear action on a higher-dimensional space, namely the action of the aforementioned translations and inversions ([8]).

*Translations*: The translate of an element $x^\mu \in \mathbb{R}^4$ by another element $a^\mu \in \mathbb{R}^4$ takes the form $x^\mu + a^\mu$. If we assume that $a^\mu = \exp(\varepsilon^\mu)$ then the fundamental vector field on $\mathbb{R}^4$ that is defined by the infinitesimal generator of a translation $\varepsilon^\mu$ is:

$$X = \varepsilon^\mu \frac{\partial}{\partial x^\mu}. \tag{3.25}$$

*Inversions:* If one has a hypersurface $S$ in $\mathbb{R}^n$ that is defined by an equation of the form $\phi(x^\mu) = 0$ then the *inversion* of $\mathbb{R}^n - S$ through the hypersurface $S$ is the transformation:

$$\mathbb{R}^n - S \to \mathbb{R}^n - S, \qquad x^\mu \mapsto \bar{x}^\mu = \frac{1}{\phi(x^\mu)} x^\mu. \tag{3.26}$$

Particular examples of such hypersurfaces are the unit sphere, for which $\phi(x^\mu) = x^2 - 1$, and the "hyperplane at infinity" that is defined by a 1-form $\phi_\mu$, so $\phi(x^\mu) = \phi_\mu x^\mu$. The latter hyperplane was used in projective relativity to account for the electromagnetic potential 1-form in the context of projective geometry; a choice of electromagnetic gauge was then equivalent to a choice of hyperplane at infinity.

The $n$ inversions of the canonical basis vectors generate an abelian group that is isomorphic to the translation group $\mathbb{R}^n$. For the linear case, $\phi(x^\mu) = \phi_\mu x^\mu$, the group parameters are defined by the components of $\phi$ relative to a choice of frame on $\mathbb{R}^n$. Hence, in order to find the form of the infinitesimal generator of an inversion, one defines a differentiable curve $x^\mu(s)$ with $dx^\mu/ds|_{s=0} \equiv v^\mu$, and then differentiates the curve at $s = 0$:

$$X^\mu = \frac{d}{ds}\bigg|_{s=0} \left( \frac{1}{\phi_\lambda x^\lambda(s)} x^\mu(s) \right) = \frac{-1}{(\phi_\lambda x^\lambda)^2} [(\phi_\lambda v^\lambda) x^\mu - (\phi_\lambda x^\lambda) v^\mu ]. \tag{3.27}$$

If one computes the differential of the inversion transformation:

$$\frac{\partial}{\partial x^\nu} \left( \frac{1}{\phi_\lambda x^\lambda} x^\mu \right) = \frac{-1}{(\phi_\lambda x^\lambda)^2} [\phi_\lambda \delta_\nu^\lambda x^\mu - (\phi_\lambda x^\lambda) \delta_\nu^\mu ] \tag{3.28}$$

---

[8] Indeed, there is a general theorem in transformation group theory that says that *any* group action on a differentiable manifold can be equivariantly represented by a linear action of that group on a vector space of sufficiently high dimension.

one can also obtain (3.27) from the chain rule upon multiplying by $v^\nu$.

One generally regards the inversions as elements of some larger group, such as the conformal Lorentz group, or the group of projective transformations of $\mathbb{RP}^n$, which we shall discuss shortly. For the moment, we shall examine the nature of the infinitesimal generator of an inversion of $\mathbb{R}^4 - \mathscr{C}$ through the light cone $\mathscr{C}$ in Minkowski space and what sort of fundamental vector field it corresponds to on $\mathbb{R}^4$.

In the case of Minkowski space; i.e., $n = 4$ and $\phi(x^\mu) = x^2 = \eta_{\mu\nu} x^\mu x^\nu$, the hypersurface at infinity is the light cone. The matrix of the differential of the inversion transformation in this case becomes:

$$\frac{\partial}{\partial x^\nu}\left(\frac{1}{\eta_{\kappa\lambda}x^\kappa x^\lambda} x^\mu\right) = -\frac{1}{x^4}(2x^\mu x_\nu - x^2 \delta^\mu_\nu). \tag{3.29}$$

Since the inversion is a diffeomorphism where it is defined, the differential of the transformation is undefined on the light cone in each tangent space, but non-singular for all of the other vectors in each tangent space.

If one multiplies the right-hand side of (3.39) by $v^\nu$, as before, one obtains the infinitesimal generator of an inversion through the light cone:

$$X^\mu = -\frac{1}{x^4}[2(x_\nu v^\nu) x^\mu - x^2 v^\mu]. \tag{3.30}$$

Inversions have an important interpretation in physical mechanics: Suppose that the coordinate frame $\partial/\partial x^\mu$, which is defined by the coordinate system $(U, x^\mu)$, and the frame $\partial/\partial \bar{x}^\mu$, which is defined by the coordinate system $(V, \bar{x}^\mu)$, have a relative velocity $v^\mu(\tau) = a^\mu \tau$ with a constant acceleration $a^\mu$. Then for any points $x^\mu \in U \cap V$ that are not on a light ray through **a** the transition from the first coordinate system to the second one can be obtained from the composition of an inversion through the light cone, a translation $\mathbf{x} \mapsto \mathbf{x} - \mathbf{a}$, and another inversion through the light cone ([9]). The result is:

$$\bar{\mathbf{x}} = \frac{1}{\eta(x^{-2}\mathbf{x}-\mathbf{a}, x^{-2}\mathbf{x}-\mathbf{a})}[\frac{1}{x^2}\mathbf{x}-\mathbf{a}] = \frac{1}{1-2\eta(\mathbf{x},\mathbf{a})+x^2 a^2}[\mathbf{x}-x^2 \mathbf{a}], \tag{3.31}$$

which has essentially the same form as (3.30). One can deal with infinitesimal inversions in either form, although the form (3.31) is the one that is used most commonly in physics, but the form (3.30) will be more useful later in relating inversions to the prolongations of the linear conformal Lie algebra.

*Projective transformations:* Both of the aforementioned nonlinear actions on $\mathbb{R}^4$ – translations and inversions through hyperplanes – can also be represented by the local form of the projection of the linear action of $SL(5; \mathbb{R})$ on $\mathbb{R}^5 - 0$ to its nonlinear action on $\mathbb{RP}^4$, by way of the projection of $\mathbb{R}^5 - 0$ onto $\mathbb{RP}^4$. One simply chooses a *Plücker chart* on an open subset of $\mathbb{RP}^4$. For instance, one can start with all $(x^0, x^1, \ldots, x^4) \in \mathbb{R}^5 - 0$ for a fixed non-zero value of $x^0$ and project onto a chart on $\mathbb{RP}^4$ by way of:

$$\mathbb{R}^5 - 0 \to \mathbb{RP}^4, \quad (x^0, x^1, \ldots, x^4) \mapsto (X^1, \ldots, X^4) = \left(\frac{x^1}{x^0}, \ldots, \frac{x^4}{x^0}\right). \tag{3.32}$$

---

[9] This relationship was first pointed by Engstrom and Zorn [**12**].

The coordinates $x^\mu$, $\mu = 0, \ldots, 4$ are called the *homogeneous* coordinates of a line in $\mathbb{RP}^4$ and the $X^i$, $i = 1, \ldots, 4$ are called its *inhomogeneous* coordinates.

The linear action of $A^\mu_\nu \in SL(5; \mathbb{R})$ on $\mathbb{R}^5 - 0$ projects to a nonlinear action on $\mathbb{RP}^4$ by way of *linear fractional transformations:*

$$X^i \mapsto \frac{A^i_0 + A^i_j X^j}{A^0_0 + A^0_j X^j}. \tag{3.33}$$

If one exhibits the matrix $A^\mu_\nu$ in partitioned form as:

$$A^\mu_\nu = \left[\begin{array}{c|c} A^0_0 & A^0_j \\ \hline A^i_0 & A^i_j \end{array}\right] \tag{3.34}$$

then one can identify the matrices of $SL(5; \mathbb{R})$ that correspond to homotheties of $\mathbb{R}^4$, translations of $\mathbb{R}^4$, inversions of $\mathbb{R}^4$ through the affine hyperplane defined by $A^0_j x^j = -1$, and invertible linear maps of $\mathbb{R}^4$ as the matrices of the form:

$$\left[\begin{array}{c|c} A^0_0 & 0 \\ \hline 0 & \delta^i_j \end{array}\right], \quad \left[\begin{array}{c|c} 1 & 0 \\ \hline A^i_0 & \delta^i_j \end{array}\right], \quad \left[\begin{array}{c|c} 1 & A^0_j \\ \hline 0 & \delta^i_j \end{array}\right], \quad \left[\begin{array}{c|c} 1 & 0 \\ \hline 0 & A^i_j \end{array}\right], \tag{3.35}$$

resp.

It will be useful for us later on to know what form the infinitesimal generator of an inversion through an affine hyperplane will take as a vector field on $\mathbb{R}^4$. First, we define a differentiable curve through the identity in $SL(5; \mathbb{R})$:

$$\mathcal{I}(s) = \left[\begin{array}{c|c} 1 & A_\lambda(s) \\ \hline 0 & \delta^\mu_\lambda \end{array}\right] \tag{3.36}$$

in which $A_\lambda(0) = 0$ for all $\lambda$ and we define:

$$a_\lambda \equiv \left.\frac{dA_\lambda}{ds}\right|_{s=0}. \tag{3.37}$$

If we differentiate the action of this one-parameter subgroup on $x^\mu$ by linear fractional transformations:

$$x^\mu(s) = \frac{1}{1 + A_\lambda(s) x^\lambda} x^\mu \qquad (3.38)$$

then we get:

$$X^\mu = \frac{dx^\mu}{ds}\bigg|_{s=0} = -(a_\lambda x^\lambda) x^\mu. \qquad (3.39)$$

One should contrast this with the form (3.27) of the infinitesimal generator of an inversion through a hyperplane that we obtained in the previous section.

Hence, a local vector field $X$ on $\mathbb{R}^4$ that represents the infinitesimal generator of an inversion through the affine hyperplane $a_k x^k = -1$ will take the form:

$$X = (a_\lambda x^\lambda) x^\mu \frac{\partial}{\partial x^\mu}, \qquad (3.40)$$

in which we have absorbed the sign into the definition of $a_\nu$.

### 3.2.2  Fiber transformations

The transformations of the fiber $\Lambda^2(\mathbb{R}^4)$ include scalar multiplication and translation, which it inherits from its vector space structure, as well as the duality isomorphism:

$$*: \Lambda^2(\mathbb{R}^4) \to \Lambda^2(\mathbb{R}^4), \qquad F_{\mu\nu} \mapsto *^{\alpha\beta}_{\mu\nu} F_{\alpha\beta} \qquad (3.41)$$

that results from imposing a complex structure * on $\Lambda^2(\mathbb{R}^4)$. We shall see that this latter assumption represents the pre-metric generalization of introducing a Lorentzian metric in electromagnetic theory, and when one already has a unit-volume element on $T(M)$, it is equivalent to defining a linear electromagnetic constitutive law.

The infinitesimal generators of the scalar multiplication of a 2-form $F_{\mu\nu} \in \Lambda^2(\mathbb{R}^4)$, its translation by another 2-form, and its duality transform define fundamental vector fields on $\Lambda^2(\mathbb{R}^4)$ that take the forms:

$$\lambda F_{\mu\nu} \frac{\partial}{\partial F_{\mu\nu}}, \qquad \phi_{\mu\nu} \frac{\partial}{\partial F_{\mu\nu}}, \qquad *^{\alpha\beta}_{\mu\nu} F_{\alpha\beta} \frac{\partial}{\partial F_{\mu\nu}}, \qquad (3.42)$$

resp.

*Linear transformations*:  Since any fiber of $\Lambda^2(\mathbb{R}^4)$ is a real six-dimensional vector space, it is naturally acted on by the group $GL(6; \mathbb{R})$ when one chooses a frame. Indeed, the associated principal fiber bundle to $\Lambda^2(\mathbb{R}^4)$, namely, the bundle of linear frames in the fibers of $\Lambda^2(\mathbb{R}^4)$, is going to be a $GL(6; \mathbb{R})$-principal bundle.

In order to represent the elements of $GL(6; \mathbb{R})$ by matrices relative to a choice of frame on the fiber $\Lambda^2(\mathbb{R}^4) = \mathbb{R}^{4*} \wedge \mathbb{R}^{4*}$, it is more convenient to denote the members of that frame by a single index that ranges from 1 to 6: $E^I$, $I = 1, \ldots, 6$. If we choose $E^i = dx^0$

$\wedge\, dx^i$ for $i = 1, 2, 3$ and $E^{i+3} = 1/2\, \varepsilon_{ijk}\, dx^j \wedge dx^k$ for $i, j, k = 1, 2, 3$ then a typical 2-form takes the form:

$$F = F_J\, E^J = \tfrac{1}{2} F_{\mu\nu}\, dx^\mu \wedge dx^\nu\,. \tag{3.43}$$

With such a choice of frame notation, an element of $GL(6; \mathbb{R})$ can be represented by a real $6\times 6$ matrix in the form $A_J^I$. We can then represent the infinitesimal generator of such a transformation of the fibers of $\Lambda^2(\mathbb{R}^4)$ in the form:

$$X(F) = A_I^J\, F_J\, \frac{\partial}{\partial F_I}\,. \tag{3.44}$$

Once again, any infinitesimal generators for the transformations of a subgroup of $GL(6; \mathbb{R})$ will take this form, with suitable restrictions on the form of $A_J^I$.

Nevertheless, the group $GL(6; \mathbb{R})$ is not necessarily the most interesting one for the purposes of pre-metric electromagnetism, since it is largely indifferent to the specific nature of 2-forms. For one thing, since 2-forms are second-rank covariant tensor fields on $\mathbb{R}^4$, one is often more concerned with the subgroup of $GL(6; \mathbb{R})$ onto which $GL(4; \mathbb{R})$ gets represented by its action on tensor products of vectors in $\mathbb{R}^4$:

$$(A \circ F)(\mathbf{v}, \mathbf{w}) = F(A(\mathbf{v}), A(\mathbf{w}))\,. \tag{3.45}$$

A structure that one must account for in pre-metric electromagnetism that plays a fundamental role is the unit-volume element $\mathcal{V}$ on $T(\mathbb{R}^4)$:

$$\mathcal{V} = \tfrac{1}{4!}\, \varepsilon_{\kappa\lambda\mu\nu}\, dx^\kappa \wedge dx^\lambda \wedge dx^\mu \wedge dx^\nu \tag{3.46}$$

along with its reciprocal unit-volume element $V$ on $T^*(\mathbb{R}^4)$:

$$V = \tfrac{1}{4!}\, \varepsilon^{\kappa\lambda\mu\nu}\, \partial_\kappa \wedge \partial_\lambda \wedge \partial_\mu \wedge \partial_\nu\,, \tag{3.47}$$

which, by definition, satisfies $\mathcal{V}(V) = 1$.

The effect of $V$ on $\Lambda^2(\mathbb{R}^4)$ is immediately felt by the fact that it defines a volume element on $\Lambda^2(\mathbb{R}^4)$. This is because the fiber, being a vector space, is orientable, so it has a unit-volume element in the form of a global non-vanishing 6-form:

$$V_\Lambda = \tfrac{1}{6!}\, \varepsilon^{IJ\ldots K}\, \partial_I \wedge \partial_J \wedge \ldots \wedge \partial_K\,, \tag{3.48}$$

in which we have employed the aforementioned index notation. The unit-volume element on $\Lambda^2(\mathbb{R}^4)$ is then simply the global non-vanishing 10-form $V \wedge V_\Lambda$.

This unit-volume element $V_\Lambda$ on any fiber allows us to reduce from $GL(6; \mathbb{R})$ to $SL(6; \mathbb{R})$. The infinitesimal generator of such a transformation will take the form of (3.44) with the restriction that $\mathrm{Tr}(A_J^I) = A_I^I = 0$.

The unit-volume element $V$ defines a scalar product on $\Lambda^2(\mathbb{R}^4)$ by way of:

$$\langle F, G \rangle = V(F \wedge G) = \varepsilon^{\kappa\lambda\mu\nu} F_{\kappa\lambda} G_{\mu\nu}. \tag{3.49}$$

The signature type of this scalar product is $(-1, -1, -1, +1, +1, +1)$, so the orthogonal group that corresponds to it is $O(3, 3)$, which can be reduced to its subgroup $SO(3, 3)$ of orthogonal transformations that also preserve $V \wedge V_\Lambda$. If we define $\eta_{IJ} = \mathrm{diag}(-1, -1, -1, +1, +1, +1)$ then an infinitesimal generator of a transformation in this subgroup takes the form (3.44) with the restriction that:

$$0 = A_{IJ} + A_{JI} = \eta_{IK} A_J^K + A_I^K \eta_{KJ}. \tag{3.50}$$

When one imposes the almost-complex structure * on the fibers of $\Lambda^2(\mathbb{R}^4)$, one must consider the nature of the invertible linear transformations of those fibers that preserve that almost-complex structure – i.e., which commute with * – and this defines a subgroup of $GL(6; \mathbb{R})$ that is isomorphic to $GL(3; \mathbb{C})$. For a frame in which * has the matrix:

$$*_J^I = \left[\begin{array}{c|c} 0 & -I \\ \hline I & 0 \end{array}\right] \tag{3.51}$$

a typical element of the $GL(3; \mathbb{C})$ subgroup of $GL(6; \mathbb{R})$ has the form:

$$\left[\begin{array}{c|c} A & -B \\ \hline B & A \end{array}\right] = \left[\begin{array}{c|c} A & 0 \\ \hline 0 & A \end{array}\right] + \left[\begin{array}{c|c} 0 & -B \\ \hline B & 0 \end{array}\right] = \left[\begin{array}{c|c} A & 0 \\ \hline 0 & A \end{array}\right] + * \left[\begin{array}{c|c} B & 0 \\ \hline 0 & B \end{array}\right]. \tag{3.52}$$

Clearly, this 6×6 real matrix corresponds to the 3×3 complex matrix $A + iB$. Hence, an infinitesimal generator of a transformation from $GL(3; \mathbb{C})$ can also be represented in the form (3.44), except that the indices go from 1 to 3 and the matrix $A_J^I$, as well as the covector components $F_I$, are complex. All one must do in order to be consistent with the almost-complex structure is to set:

$$iE^I = *E^I. \tag{3.53}$$

The combination of the volume element $V$ and the almost-complex structure * allows us to define another scalar product on $\Lambda^2(\mathbb{R}^4)$ by way of:

$$(F, G) = V(F \wedge *G). \tag{3.54}$$

Since the signature type of this scalar product is Euclidian, this allows us to reduce the subgroup $GL(3; \mathbb{C})$ further to $O(3; \mathbb{C})$, and then to $SO(3; \mathbb{C})$ by way of the unit-volume element $V_\Lambda$. An infinitesimal generator of such a transformation will take the form of (3.44) with the restriction that:

$$0 = A_{IJ} + A_{JI} = \delta_{IK} A_J^K + A_I^K \delta_{KJ}, \qquad *A = A*. \tag{3.55}$$

It is worth noting that even though we have yet to explicitly introduce a Lorentzian structure on the tangent bundle to our space, we have nevertheless arrived at a group that is locally isomorphic to the Lorentz group. This is the essence of the transition from metric to pre-metric electromagnetism, that the combination of a volume element and an almost-complex structure on the bundle of 2-forms gives one a Lorentzian geometry as a special case of a more general geometric framework, in which the geometry of spacetime is fundamentally determined by its electromagnetic constitutive properties ([10]).

### 3.2.3 Contact transformations

By analogy with the case of the canonical 2-form $d\theta$ on $T^*(M)$ or $J^1(\mathbb{R}, M)$, we shall refer to the fiber-preserving diffeomorphisms $f$ of $\Lambda^2(M)$ that preserve the canonical 2-form $d\Phi$ on $\Lambda^2(M)$, up to a scalar factor – i.e.:

$$f^*d\Phi = \alpha\, d\Phi, \tag{3.56}$$

as *contact transformations*; in this expression, $\alpha$ is some function on $\Lambda^2(M)$. These transformations will be symmetries of the exterior differential system – i.e., the ideal – that $d\Phi$ generates, which will essentially be the exterior differential equation $d\Phi = 0$.

The infinitesimal generators of contact transformations will then be vector fields $X$ on $\Lambda^2(M)$ that satisfy:

$$L_X d\Phi = d i_X d\Phi = \alpha\, d\Phi. \tag{3.57}$$

Locally, this takes the form:

$$\tfrac{1}{2} X_{\mu\nu,\lambda}\, dx^\lambda \wedge dx^\mu \wedge dx^\nu + \tfrac{1}{2}(X_{\mu\nu}{}^{,\kappa\lambda} + X^\kappa{}_{,\mu}\, \delta^\lambda{}_\nu)\, dF_{\kappa\lambda} \wedge dx^\mu \wedge dx^\nu$$
$$+ X^{\mu,\kappa\lambda}\, dF_{\kappa\lambda} \wedge dF_{\mu\nu} \wedge dx^\nu = \tfrac{1}{2}\alpha\, \delta^\kappa{}_\mu \delta^\lambda{}_\nu\, dF_{\kappa\lambda} \wedge dx^\mu \wedge dx^\nu, \tag{3.58}$$

which then gives the following system of partial differential equations for the components of $X$:

$$\frac{\partial X_{\mu\nu}}{\partial x^\lambda} = \frac{\partial X^\mu}{\partial F_{\mu\nu}} = 0, \qquad \frac{\partial X^\kappa}{\partial x^\mu}\delta^\lambda{}_\nu + \tfrac{1}{2}\frac{\partial X_{\mu\nu}}{\partial F_{\kappa\lambda}} = \tfrac{1}{2}\alpha\, \delta^\kappa{}_\mu \delta^\lambda{}_\nu. \tag{3.59}$$

The first two equations tell us that $X^\mu = X^\mu(x^\lambda)$, $X_{\mu\nu} = X_{\mu\nu}(F_{\kappa\lambda})$. The second breaks into:

$$\frac{\partial X^\kappa}{\partial x^\mu} = \alpha_1 \delta^\kappa{}_\mu, \qquad \frac{\partial X_{\mu\nu}}{\partial F_{\kappa\lambda}} = \alpha_2\, \delta^\kappa{}_\mu \delta^\lambda{}_\nu, \tag{3.60}$$

in which $\alpha_1$, $\alpha_2$ are new functions on $\Lambda^2(\mathbb{R}^4)$.

The solutions to these equations take the form:

---

[10] For more details on the role of complex geometry in pre-metric electromagnetism, cf. Delphenich [**13**].

$$X = (\varepsilon^\mu + \alpha_1 x^\mu)\frac{\partial}{\partial x^\mu} + (\varepsilon_{\mu\nu} + \alpha_2 F_{\mu\nu})\frac{\partial}{\partial F_{\kappa\lambda}}, \qquad (3.61)$$

which generates translations and dilatations of the coordinates of the points of $\mathbb{R}^4$ and the components of the 2-form *F*. One must expect that these transformations would follow from the linearity of the characteristic system as a system of PDE's.

Eventually, we shall include the canonical 3-form $d\Phi$ as a generator of our eventual system of pre-metric electromagnetic field equations. However, we shall expand the list of possible symmetries to $d\Phi$ by allowing $L_X d\Phi$ to take its values in a larger ideal.

## 4  Mathematical generalizations

The computations that we were making in the previous section admit some useful mathematical generalizations that we shall apply in our study of the symmetries of pre-metric electromagnetism, namely, Lie equations and the prolongations of Lie algebras.

### 4.1  Lie equations

One of the pillars of Lie's theory of transformation groups was his third theorem, which was proved by É. Cartan, and which addressed the question of whether any Lie algebra that could be defined abstractly over a vector space, such as $\mathbb{R}^n$, could be represented by the Lie algebra of left-invariant vector fields on some Lie group. This is really a problem of the integration of a system of partial differential equations – viz., the Maurer-Cartan equations:

$$d\theta^i = -\tfrac{1}{2} c^i_{jk} [\theta^j, \theta^k] \qquad (4.1)$$

when one is given the structure constants $c^i_{jk}$ of the Lie algebra. The solution $\theta^i$ to this system is a left-invariant coframe field on the identity component of a Lie group with a left-invariant frame field $X^i$ that is reciprocal to $\theta^i$ and satisfies the defining equation for the Lie algebra in question:

$$[X^i, X^j] = c^k_{ij} X^k. \qquad (4.2)$$

In fact, a direct application of the intrinsic formula for the exterior derivative:

$$d\theta(X, Y) = X\theta(Y) - Y\theta(X) - \theta([X, Y]) \qquad (4.3)$$

shows that the system (4.2) is equivalent to the system (4.1).

More generally, one characterizes a *Lie equation* ([11]) as a system of partial differential equations on a manifold whose solution is a vector field that generates a one-parameter subgroup of transformations that are obtained from a given Lie group.

---

[11] The definition that we shall use for a Lie equation is considerably more specialized than the one that is given by Kumpera and Spencer [**14**], Malgrange [**15**], Goldschmidt [**16**], and others, although their definition essentially prolongs our definition to a submanifold of a bundle of *k*-jets. That is because their

The equation that serves as role model for a broad class of such equations is the *Killing equation:*

$$L_X g = 0 , \qquad (4.4)$$

on a Riemannian or pseudo-Riemannian manifold $(M, g)$. Its solution $X$ is a vector field on $M$ that serves as the infinitesimal generator of a one-parameter subgroup of isometries of $g$. Hence, if $O(p, q)$ is the orthogonal group that is associated with $g$ and $\mathfrak{so}(p, q)$ is its Lie algebra then one expects that there should be an element $\omega \in \mathfrak{so}(p, q)$ that makes $X$ the fundamental vector field of some action of $\exp(\omega s)$ on $M$ – i.e., some representation of $\exp(\omega s)$ in the group $\text{Diff}(M)$ of diffeomorphisms of $M$. By differentiation at the identity, this will associate $\omega \in \mathfrak{so}(p, q)$ with the vector field $X \in \mathfrak{X}(M)$.

Note that one cannot expect a representation of the entire group $O(p, q)$ in $\text{Diff}(M)$ since many (pseudo-)Riemannian manifolds admit *no* Killing vector fields, much less one for every generator of $\mathfrak{so}(p, q)$.

One can put (4.4) into a different form, by making use of Killing's equation:

$$(L_X g)_{\mu\nu} = X_{\mu;\nu} + X_{\nu;\mu} , \qquad (4.5)$$

in which $X_\mu = g_{\mu\nu} X^\nu$ are components of the 1-form $i_X g$ that is metric-dual to the vector field $X$, and the semi-colon refers to covariant differentiation using the Levi-Civita connection that is associated with $g$.

In the case where $(M, g)$ is Minkowski space $\mathfrak{M}_4 = (\mathbb{R}^4, \eta)$ (4.5) takes the form:

$$(L_X \eta)_{\mu\nu} = \frac{\partial X_\mu}{\partial x^\nu} + \frac{\partial X_\nu}{\partial x^\mu} \qquad (4.6)$$

relative to the natural frame $\partial_\mu$ on $\mathbb{R}^4$.

Now, one notes that the right-hand side of (4.6) is twice the symmetric part $\sigma_{\mu\nu}$ of $\partial X_\mu / \partial x^\nu$ when one polarizes it in the usual way ([12]):

$$\frac{\partial X_\mu}{\partial x^\nu} = \sigma_{\mu\nu} + \omega_{\mu\nu} = \tfrac{1}{2}\left(\frac{\partial X_\mu}{\partial x^\nu} + \frac{\partial X_\nu}{\partial x^\mu}\right) + \tfrac{1}{2}\left(\frac{\partial X_\mu}{\partial x^\nu} - \frac{\partial X_\nu}{\partial x^\mu}\right). \qquad (4.7)$$

Hence, one can say that the Killing equation is equivalent to the system of linear first-order partial differential equations:

---

objective was examining the formal integrability and deformation of Lie equations, whereas ours is simply to apply the basic notions of the theory of Lie equations to the problem of symmetries of differential system on manifolds.

[12] In continuum mechanics, if $X$ is a displacement vector field then $\sigma$ is the infinitesimal strain and $\omega$ is the infinitesimal rotation that it defines. If $X$ were a flow velocity vector field then $\sigma$ would be its infinitesimal rate of strain and $\omega$ would be its vorticity.

$$\frac{\partial X_\mu}{\partial x^\nu} = \omega_{\mu\nu}, \tag{4.8}$$

or, more to the point:

$$\frac{\partial X^\mu}{\partial x^\nu} = \omega^\mu{}_\nu, \tag{4.9}$$

in which we have raised the index $\mu$ using $\eta_{\mu\nu}$. Since $\omega^\mu{}_\nu \in \mathfrak{so}(3, 1)$, we can also see that (4.9) is the defining equation for the fundamental vector field on $\mathbb{R}^4$ that $\omega^\mu{}_\nu$ generates by way of its defining representation in $\mathfrak{gl}(4; \mathbb{R})$.

As long as our manifold in question is $\mathfrak{M}_4$, (4.9) can be integrated easily to give us:

$$X^\mu = \varepsilon^\mu + \omega^\mu{}_\nu x^\nu, \tag{4.10}$$

which shows that we can faithfully represent the Poincaré Lie algebra in $\mathfrak{X}(\mathbb{R}^4)$ by vector fields of the form:

$$X = \varepsilon^\mu \frac{\partial}{\partial x^\mu}, \qquad X = \omega^\mu{}_\nu x^\nu \frac{\partial}{\partial x^\mu}, \tag{4.11}$$

which represent infinitesimal translations and infinitesimal Lorentz transformations, respectively.

More generally, when a reduction of the bundle $GL(M)$ of linear frames on a manifold $M$ to a bundle of linear frames that all lie in an orbit of the action of a subgroup $G \subset GL(n)$ – i.e., a *G-structure* ([13]) – is defined by a fundamental tensor field $t$ the Lie equation that is associated with the action of $G$ on either $M$ or $GL(M)$ is the following generalization of the Killing equation:

$$L_X t = 0. \tag{4.12}$$

This is the equation whose prolongation to a higher-order jet bundle is usually referred to as a *Lie equation*, but we shall use that term here for equations of the form (4.12) themselves. One finds that for vector fields on $\mathbb{R}^n$ (4.12) is generally equivalent to an equation of the form:

$$\frac{\partial X^\mu}{\partial x^\nu} = \alpha^\mu{}_\nu, \tag{4.13}$$

where $\alpha^\mu{}_\nu \in \mathfrak{g}$, the Lie algebra of $G$. As long as $\alpha^\mu{}_\nu$ is a constant matrix, this equation integrates to the vector field:

$$X = (\varepsilon^\mu + \alpha^\mu{}_\nu x^\nu) \frac{\partial}{\partial x^\mu}, \tag{4.14}$$

---

[13] For a more detailed discussion of the application of the theory of *G*-structures to the spacetime manifold, see [**17**], and confer the references cited therein for some of the key references on the mathematics of *G*-structures.

in which the constants $\varepsilon^\mu$ represent either the infinitesimal generators of a translation or the integration constants.

In order to deal with the case in which the right-hand side of (4.13) is not constant, but a function of $x^\lambda$, one must go to a "prolongation" of $\mathfrak{g}$, which we will discuss in due course, but first, we make a few more comments concerning Lie equations.

When the symmetry group is $SL(4; \mathbb{R})$, the fundamental tensor field that is associated with the reduction from the bundle of linear frames to the bundle of oriented unit-volume frames is the unit volume element $\mathcal{V}$. Hence, the Lie equation is:

$$L_X \mathcal{V} = \frac{\partial X^\lambda}{\partial x^\lambda} \mathcal{V} = 0, \tag{4.15}$$

which is equivalent to the statement that the infinitesimal generators of volume-preserving diffeomorphisms are divergenceless vector fields. The fundamental vector field that is generated by $\alpha^\mu{}_\nu \in \mathfrak{sl}(4; \mathbb{R})$ is then of the form (4.14), with the restriction that $\alpha^\mu{}_\mu = 0$.

One might wish to generalize the form of the Lie equation to include the *conformal Killing equation*:

$$L_X g = \lambda g, \tag{4.16}$$

in which $\lambda \in C^\infty(M)$.

Suppose that we are dealing with Minkowski space, as before, so $g = \eta$. Since the left-hand side of (4.16) is the twice the symmetric part of $\partial X_\mu / \partial x^\nu$, we see that (4.16) is associated with a system of equations of the form:

$$\frac{\partial X_\mu}{\partial x^\nu} = \omega_{\mu\nu} + \lambda \, \eta_{\mu\nu}, \tag{4.17}$$

or:

$$\frac{\partial X^\mu}{\partial x^\nu} = \omega^\mu{}_\nu + \lambda \, \delta^\mu{}_\nu. \tag{4.18}$$

When $\lambda$ is a constant scalar – hence, $\omega^\mu{}_\nu$ is a constant matrix – one sees that the infinitesimal transformations that are defined by (4.18) come from the *linear* conformal Lorentz Lie algebra, $\mathfrak{co}(3, 1)$. A fundamental vector field then takes the form:

$$X = (\varepsilon^\mu + \omega^\mu{}_\nu x^\nu + \lambda x^\mu) \frac{\partial}{\partial x^\mu}, \tag{4.19}$$

which is simply the sum of the generators of an infinitesimal translation, an infinitesimal Lorentz transformation, and an infinitesimal dilatation, as expected.

However, $\lambda$ does not have to be constant, and, as we shall see next, when $\lambda$ is a linear functional on $\mathbb{R}^4$, the resulting infinitesimal transformations:

$$\frac{\partial X^\mu}{\partial x^\nu} = \omega^\mu{}_\nu(x^\lambda) + \lambda(x^\lambda) \delta^\mu{}_\nu. \tag{4.20}$$

come from the "first prolongation" of $\mathfrak{co}(3, 1)$, namely, the infinitesimal inversions through the light cone.

### 4.2 Prolongations of Lie algebras [**18-21**]

If $\mathfrak{g}$ is a Lie algebra that, for our purposes, is a sub-algebra of $\mathfrak{gl}(n; \mathbb{R}) = \text{Hom}(\mathbb{R}^n, \mathbb{R}^n)$ for some $n$ then the *first prolongation of* $\mathfrak{g}$, which is denoted by $\mathfrak{g}^{(1)}$, is defined to be the Lie sub-algebra of $\text{Hom}(\mathbb{R}^n, \mathfrak{g})$ that consists of all $T \in \text{Hom}(\mathbb{R}^n, \mathfrak{g})$ such that for all $\mathbf{v}$, $\mathbf{w} \in \mathbb{R}^n$, one has:

$$T(\mathbf{v})(\mathbf{w}) = T(\mathbf{w})(\mathbf{v}) . \tag{4.22}$$

This condition shows that $\mathfrak{g}^{(1)}$ is a vector subspace of $S^2(\mathbb{R}^n) \otimes \mathbb{R}^n$; i.e., its elements are symmetric second-rank covariant tensors on $\mathbb{R}^n$ with values in $\mathbb{R}^n$. Since we are dealing with $\mathbb{R}^n$ and $\mathfrak{gl}(n; \mathbb{R})$ in particular, we can represent an element of $\mathfrak{g}^{(1)}$ in terms of a basis $\mathbf{e}_i$, $i = 1, \ldots, n$ for $\mathbb{R}^n$ and its reciprocal basis $\theta^i$ on $\mathbb{R}^{n*}$ as:

$$T = T_{ij}^k \ \theta^i \otimes \theta^j \otimes \mathbf{e}_k \qquad (T_{ij}^k = T_{ji}^k) . \tag{4.23}$$

One can recursively define higher prolongations of $\mathfrak{g}$ by way of $\mathfrak{g}^{(k+1)} = (\mathfrak{g}^k)^{(1)}$; hence, the $k^{\text{th}}$ prolongation of $\mathfrak{g}$ will be a vector subspace of $S^k(\mathbb{R}^n) \otimes \mathbb{R}^n$. A Lie algebra $\mathfrak{g}$ will be said to be of *finite type* if there is an integer $k$ such that $\mathfrak{g}^{(l)} = 0$ for all $l \geq k$; if no such $k$ exists then $\mathfrak{g}$ is said to have *infinite type*.

In any event, one can form the infinite sum (not necessarily a *direct* sum, though; hence, we allow sums with an infinite number of non-zero terms):

$$\mathfrak{g}[[n]] = \mathbb{R}^n \oplus \mathfrak{g} \oplus \mathfrak{g}^{(1)} \oplus \ldots \oplus \mathfrak{g}^{(k)} \oplus \ldots , \tag{4.24}$$

which we call the *formal algebra* of $\mathfrak{g}$. It is analogous to the ring $\mathbb{R}^n[[x^i]]$ of formal power series in $n$ real variables that take their values in $n$ real numbers, since, when they are applied to the vector fields on $\mathbb{R}^n$ − regarded as linear differential operators – and then applied to a smooth function $f$ on $\mathbb{R}^n$ this gives one the Taylor series expansion of $f$ about a given point.

Since we only said "smooth" and not "analytic" the use of the word "formal" is to suggest that one is ignoring the issue of convergence of the series; generally, one will consider only finite series expressions, anyway. One notes that in the basis-free approach to analysis (cf., Dieudonné [**22**], Lang [**23**]), the higher-order derivatives of a function $X$: $\mathbb{R}^n \to \mathbb{R}^n$ will be elements of the successive spaces $S^k(\mathbb{R}^n) \otimes \mathbb{R}^n$. In effect, we have generalized the machinery of Taylor expansions of local vector fields on $n$-dimensional differentiable manifolds in a manner that allows one to impose symmetries upon the vector fields more directly.

In the most general case, the local equation for the infinitesimal generator $X$ of a local diffeomorphism on $\mathbb{R}^n$ – i.e., an arbitrary smooth vector field − takes the tautological form:

Symmetries and pre-metric electromagnetism                                   27$$\frac{\partial X^{\mu}}{\partial x^{\nu}} = a_{\nu}^{\mu}(x^{\lambda}) = a_{\nu}^{\mu}(x_0^{\lambda}) + a_{\nu,\alpha}^{\mu}(x_0^{\lambda})\varepsilon^{\alpha} + a_{\nu,\alpha\beta}^{\mu}(x_0^{\lambda})\varepsilon^{\alpha}\varepsilon^{\beta} + \cdots \qquad (4.25)$$

in which $x_0^{\lambda}$, $\varepsilon^{\mu} \in \mathbb{R}^n$, and the successive power series coefficients, which are essentially the successive derivatives of $X$ at $x_0$:

$$\left.\frac{\partial X^{\mu}}{\partial x^{\nu}}\right|_{x^{\mu}=x_0^{\mu}} = a_{\nu}^{\mu}(x_0^{\lambda}), \quad \ldots, \quad \left.\frac{1}{k!}\frac{\partial^k X^{\mu}}{\partial x^{\nu}\partial x^{\alpha}\cdots\partial x^{\beta}}\right|_{x^{\mu}=x_0^{\mu}} = a_{\nu,\alpha\cdots\beta}^{\mu}(x_0^{\lambda}), \ldots \qquad (4.26)$$

then define elements of the successive spaces $S^k(\mathbb{R}^n) \otimes \mathbb{R}^n$. Hence, if we want $X$ to only generate transformations in some subgroup $G$ of $GL(n; \mathbb{R})$ whose Lie algebra is $\mathfrak{g}$ then we will need to make similar restrictions to the spaces in which the coefficients of the higher-order terms of (4.25) can take their values.

We can also regard (4.26) as a sequence of linear PDE's in $X^{\mu}$ when each of the *a*-matrices is a given element of $S^k(\mathbb{R}^n) \otimes \mathbb{R}^n$. Hence, (4.25) is essentially a generalization – more precisely, a prolongation – of a Lie equation to a higher order system.

A useful generalization of the formal algebra of a Lie algebra $\mathfrak{g}$ that we shall use in the sequel is the formal algebra of a *given* finite sequence of Lie algebras $\mathfrak{g}, \mathfrak{g}^1, \ldots, \mathfrak{g}^k$ such that $\mathfrak{g}$ is a sub-algebra of $\mathfrak{gl}(n; \mathbb{R})$ and each $\mathfrak{g}^i$ is a sub-algebra of $S^{i+1}(\mathbb{R}^n) \otimes \mathbb{R}^n$:

$$\{\mathfrak{g}, \mathfrak{g}^1, \ldots, \mathfrak{g}^k\}[[n]] = \mathbb{R}^n \oplus \mathfrak{g}^1 \oplus \ldots \oplus \mathfrak{g}^k \oplus \mathfrak{g}^{k(1)} \oplus \mathfrak{g}^{k(2)} \oplus \ldots ; \qquad (4.27)$$

however, we are not specifying that each of the given Lie algebras be the prolongation of the previous one, only the $\mathfrak{g}^{k(1)}$, $\mathfrak{g}^{k(2)}$, etc. Hence, we are looking at all possible Taylor series expansions for $X$ in which we have specified the Lie algebras in which the first $k+1$ terms take their values.

In order to insure that this formal algebra represents a Lie algebra of vector fields, we need to impose the further condition ([14]):

$$[\mathfrak{g}^r, \mathfrak{g}^s] \subset \mathfrak{g}^{r+s}. \qquad (4.28)$$

Let us now apply the methods of prolongation to some of the Lie algebras that will be of concern to us later.

The Lie algebra $\mathfrak{so}(p, q)$ has finite type since the anti-symmetry of its matrices (with the upper index lowered) leads to the result that $\mathfrak{so}(p, q)^{(1)} = 0$; hence, $\mathfrak{so}(p, q)$ has type 1, and in the Lie equation for its infinitesimal generators on $\mathbb{R}^n$:

$$\frac{\partial X^{\mu}}{\partial x^{\nu}} = \omega_{\nu}^{\mu} \qquad (4.29)$$

the right-hand side can only be a constant matrix in $\mathfrak{so}(p, q)$. The solutions of (4.29) then take the form:

---

[14] For $r+s > k$, one regards $\mathfrak{g}^{r+s}$ as $(\mathfrak{g}^k)^{(r+s-k)}$.

$$X^\mu = \varepsilon^\mu + \omega^\mu_\nu x^\nu. \tag{4.30}$$

Since the formal algebra of $\mathfrak{so}(p, q)$ is simply $\mathbb{R}^n \oplus \mathfrak{so}(p, q)$, and in the case of $\mathfrak{so}(3, 1)$, this is the Lie algebra of the Poincaré group, we then see that one indeed has $X \in \mathfrak{so}(p, q)[[n]]$.

Now, let us consider the case of the linear conformal Lorentz Lie algebra $\mathfrak{co}(3, 1)$. Its first prolongation $\mathfrak{co}(3,1)^{(1)}$ is defined to be a subspace of the vector space $\mathrm{Hom}(\mathbb{R}^4, \mathfrak{co}(3, 1))$, whose elements are linear maps $T$ from $\mathbb{R}^4$ to $\mathfrak{co}(3, 1)$ that satisfy the defining condition (4.22) for a prolongation. Hence, its elements will take the form of symmetric second-rank covariant tensors on $\mathbb{R}^4$ with values in $\mathbb{R}^4$.

If we take:

$$T^\mu_{\ \nu}(\mathbf{v}) = \omega^\mu_{\ \nu}(\mathbf{v}) + \alpha(\mathbf{v})\delta^\mu_\nu, \tag{4.31}$$

to be linear in $\mathbf{v}$ then the matrix of $T$ (with the upper index lowered by means of $\eta_{\mu\nu}$) must take the form:

$$T_{\mu\nu\alpha} = \omega_{\mu\nu\alpha} + a_\alpha \eta_{\mu\nu}. \tag{4.32}$$

By exploiting the symmetries in the indices, one finds that the only possible form for $T_{\mu\nu\alpha}$ is:

$$T_{\mu\nu\alpha} = a_\nu \eta_{\mu\alpha} - a_\mu \eta_{\alpha\nu} + a_\alpha \eta_{\mu\nu}, \tag{4.33}$$

so:

$$T^\mu_{\ \alpha\nu} = a_\lambda(\delta^\lambda_\nu \delta^\mu_\alpha + \delta^\lambda_\alpha \delta^\mu_\nu - \eta^{\mu\lambda}\eta_{\alpha\nu}). \tag{4.34}$$

One has a linear isomorphism of $\mathfrak{co}(3,1)^{(1)}$ with $\mathbb{R}^{4*}$ that is defined by:

$$\alpha: \mathfrak{co}(3,1)^{(1)} \to \mathbb{R}^{4*}, \qquad T \mapsto \alpha_T = a_\mu \theta^\mu, \tag{4.35}$$

so:

$$\alpha_T(\mathbf{v}) \equiv \alpha(T(\mathbf{v})) = a_\mu v^\mu, \tag{4.36}$$

and $\alpha$ is also the linear functional $\lambda$ on $\mathfrak{co}(3,1)$ that appears in the definition (4.16) of $\mathfrak{co}(3,1)$.

One notes that the (vector-valued) quadratic form that is associated with $T$ takes the form:

$$T(\mathbf{v})(\mathbf{v}) = 2\eta(\mathbf{a}, \mathbf{v})\mathbf{v} - \eta(\mathbf{v}, \mathbf{v})\mathbf{a}, \tag{4.37}$$

in which the vector $\mathbf{a}$ is metric-dual to the covector $\alpha$. Up to a scalar factor, this is the same form that the infinitesimal generators of inversions through the light cone in Minkowski space have.

When one goes to the next prolongation of $\mathfrak{co}(3,1)$, one finds that $\mathfrak{co}(3,1)^{(2)} = 0$, so $\mathfrak{co}(3, 1)$ is of type 2. However, one should point out the $\mathfrak{co}(2)$ is of infinite type; indeed, it is only for dimension 2 that the conformal Lie algebra is of infinite type.

Ultimately, we find that the full Lie algebra of the conformal Lorentz group, which we defined as a sub-algebra of $\mathfrak{X}(\mathbb{R}^4)$, is also the formal algebra:

$$\mathfrak{co}(3,1)[[4]] = \mathbb{R}^4 \oplus \mathfrak{co}(3,1) \oplus \mathfrak{co}(3,1)^{(1)},$$

which consists of infinitesimal translations, homotheties, Lorentz transformations, and inversions through the light cone, respectively. Hence, the most general solution to the Lie equation:

$$\frac{\partial X^\mu}{\partial x^\nu} = \omega^\mu_\nu(x^\lambda) + \alpha(x^\lambda)\delta^\mu_\nu, \tag{4.38}$$

that belongs to $\mathfrak{co}(3,1)[[4]]$ is:

$$X^\mu = \varepsilon^\mu + \omega^\mu_\nu x^\nu + \alpha x^\nu + 2 x_\nu a^\nu x^\mu - x^2 a^\mu. \tag{4.39}$$

Note that even though it is conceivable that the system of PDE's (4.38) might have solutions for more complicated forms for the functions $\omega^\mu_\nu(x^\lambda)$ and $\alpha(x^\lambda)$ than linear ones, the fact that $\mathfrak{co}(3,1)^{(2)} = 0$ truncates the Taylor series after the linear terms implies that any vector field obtained from more general equations would no longer represent the fundamental vector field of any one-parameter subgroup of Diff($M$).

Actually, just as one can find a faithful linear representation of the $n$-dimensional affine group in the $GL(n+1; \mathbb{R})$ by way of homogeneous coordinates for $\mathbb{R}P^n$, one can find a faithful linear representation of the $n$-dimensional conformal group for a particular orthogonal structure in an orthogonal group for a space of dimension $n + 2$ by the use of "polyspherical coordinates" for the hyperspheres in $S^n$. One can still represent the inversion through the light cone in Minkowski space in $SL(5; \mathbb{R})$ by matrices of the form:

$$\begin{bmatrix} 0 & x_\nu \\ \hline 0 & \delta^\mu_\nu \end{bmatrix}, \tag{4.40}$$

but the representation is still not linear, since the matrices are a function of position in $\mathbb{R}^4$.

We mention this only in passing as we shall not have occasion to apply this in the sequel.

The Lie algebras $\mathfrak{gl}(4)$ and $\tilde{\mathfrak{sl}}(4)$ are of infinite type, as one might expect from the infinite-dimensionality of the group of the diffeomorphisms of $\mathbb{R}^4$ and the group of volume-preserving diffeomorphisms of $\mathbb{R}^4$, when it is given a choice of unit-volume element. However (cf., Singer and Sternberg [**19**]), there are only four possible symmetry Lie algebras whose formal algebras start with $\mathfrak{gl}(4)$:

  *i.* $\mathfrak{gl}[[4]]$,
  *ii.* $\{\mathfrak{gl}(4), \tilde{\mathfrak{sl}}(4)^{(1)}\}[[4]]$,
  *iii.* $\{\mathfrak{gl}(4), \mathfrak{p}^{(1)}\}[[4]]$, in which we will define the Lie algebra $\mathfrak{p}^{(1)}$ shortly,
  *iv.* $\{\mathfrak{gl}(4), 0\}[[4]]$.

One notes that $\mathfrak{gl}(4)^{(k)}$ is all of $S^k(\mathbb{R}^4) \otimes \mathbb{R}^4$ for every $k$. The formal algebra $\mathfrak{gl}[[4]]$ then corresponds to allowing any smooth local vector field on $\mathbb{R}^4$ as a symmetry generator; this amounts to admitting all local diffeomorphisms as symmetries.

Now, let us investigate the formal algebra:

$$\{\mathfrak{gl}(4), \mathfrak{sl}(4)^{(1)}\}[[4]] = \mathbb{R}^4 \oplus \mathfrak{gl}(4) \oplus \mathfrak{sl}(4)^{(1)} \oplus \mathfrak{sl}(4)^{(2)} \oplus \ldots$$

The sum continues indefinitely since $\mathfrak{sl}(4)$ is of infinite type. Furthermore, since $\mathfrak{gl}(4) = \mathbb{R} \oplus \mathfrak{sl}(4)$ this formal algebra differs from $\mathfrak{sl}[[4]]$ only by the extra $\mathbb{R}$ summand, which represents a trace that one might add to a (traceless) element of $\mathfrak{sl}(4)$. Hence, just as the formal algebra $\mathfrak{sl}[[4]]$ would represent successive derivatives of vector fields on $\mathbb{R}^4$ with *zero* divergence, the formal algebra $\{\mathfrak{gl}(4), \mathfrak{sl}(4)^{(1)}\}[[4]]$ only extends this to vector fields with *constant* divergence. Hence, such vector fields should obey the Lie equation:

$$\frac{\partial X^\mu}{\partial x^\nu} = a^\mu{}_\nu(x^\lambda) + \alpha \delta^\mu{}_\nu, \tag{4.41}$$

in which $a^\mu{}_\nu(x^\lambda) \in \mathfrak{sl}(4)$ – so $a^\mu{}_\mu(x^\lambda) = 0$ – for all $x^\lambda \in \mathbb{R}^4$ and $\alpha$ is a constant. Hence, the general solution to (4.39) will take the form:

$$X^\mu = J^\mu(x^\lambda) + \alpha x^\mu, \tag{4.42}$$

in which the vector field $J$ has vanishing divergence.

By definition, the Lie algebra $\mathfrak{sl}(4)^{(1)}$ consists of all linear maps $T: \mathbb{R}^4 \to \mathfrak{sl}(4)$, such that $T(\mathbf{v})(\mathbf{w}) = T(\mathbf{w})(\mathbf{v})$ for all $\mathbf{v}, \mathbf{w} \in \mathbb{R}^4$ and $\mathrm{Tr}(T(\mathbf{v})) = 0$ for any $\mathbf{v} \in \mathbb{R}^4$. However, one cannot generally be more specific about the form an element in $\mathfrak{sl}(4)^{(1)}$ since it, like $\mathfrak{sl}(4)$ itself, admits an infinitude of prolongations.

Under the action of $\mathfrak{gl}(4)$, $S^2(\mathbb{R}^4) \otimes \mathbb{R}^4$ is reducible to $\mathfrak{sl}(4)^{(1)} \oplus \mathfrak{p}^{(1)}$, in which the action of $\mathfrak{gl}(4)$ on both direct summands is invariant. In order to explicitly construct the complementary vector space $\mathfrak{p}^{(1)}$, we let $\sigma$ denote the symmetrization map:

$$\sigma: T^{0,3}(\mathbb{R}^4) \to S^2(\mathbb{R}^4) \otimes \mathbb{R}^4, \qquad \alpha \otimes \beta \otimes \mathbf{w} \mapsto (\alpha \otimes \beta + \beta \otimes \alpha) \otimes \mathbf{w}. \tag{4.43}$$

Now, define the linear map:

$$\gamma: \mathbb{R}^{4*} \to S^2(\mathbb{R}^4) \otimes \mathbb{R}^4, \qquad \alpha \mapsto \sigma(I \otimes \alpha), \tag{4.44}$$

in which $I$ is the identity map, so we define:

$$T(\mathbf{v}) \equiv \gamma(\alpha)(\mathbf{v}) = \mathbf{v} \otimes \alpha + \alpha(\mathbf{v}) I. \tag{4.45}$$

This takes the component form:

$$T^\mu_\nu(\mathbf{v}) = (\alpha_\nu \delta^\mu{}_\lambda + \alpha_\lambda \delta^\mu{}_\nu) v^\lambda. \tag{4.46}$$

We let $\mathfrak{p}^{(1)}$ denote the image of $\mathbb{R}^{4*}$ in $S^2(\mathbb{R}^4) \otimes \mathbb{R}^4$ under $\gamma$ so we have indeed that $T: \mathbb{R}^4 \to \mathfrak{p}^{(1)}$. One notes that the quadratic form that is associated with $T$ then takes the form:

$$T(\mathbf{v})(\mathbf{v}) = 2\alpha(\mathbf{v})\mathbf{v}. \tag{4.47}$$

If one refers back to equation (4.24) then one will recognize that this expression is proportional to the expression for the infinitesimal generator of an inversion through the affine hyperplane $s\alpha(\mathbf{v}) = -1$, when $s$ is a curve parameter.

As one can verify ([15]), one has $\mathfrak{p}^{(2)} = 0$. We can then form the finite-dimensional formal algebra:

$$\{\mathfrak{gl}(4), \mathfrak{p}^{(1)}\}[[4]] = \mathbb{R}^4 \oplus \mathfrak{gl}(4) \oplus \mathfrak{p}^{(1)}.$$

Since $\mathfrak{p}^{(1)}$ is linearly isomorphic to $\mathbb{R}^{4*}$, we see that this algebra is linearly isomorphic to the Lie algebra $\mathfrak{sl}(5; \mathbb{R})$, which represents the infinitesimal generators of projective transformations of $\mathbb{R}P^4$, when its points are represented by homogeneous coordinates in $\mathbb{R}^5$ minus a hyperplane at infinity. Hence, this choice of formal algebra puts one into the realm of projective differential geometry.

The Lie equation for a vector field in this case is:

$$\frac{\partial X^\mu}{\partial x^\nu} = \alpha \delta^\mu_\nu + a^\mu_\nu + (\alpha_\nu \delta^\mu_\lambda + \alpha_\lambda \delta^\mu_\nu) x^\lambda, \tag{4.48}$$

in which the scalar $\alpha$ and the matrix $a^\mu_\nu \in \mathfrak{sl}(4)$ are constant. Hence, the solutions will be of the form:

$$X^\mu = \varepsilon^\mu + \alpha x^\mu + a^\mu_\nu x^\nu + (\alpha_\nu x^\nu) x^\mu. \tag{4.49}$$

The fourth case, in which the formal algebra is $\{\mathfrak{gl}(4), 0\}[[4]] = \mathbb{R}^4 \oplus \mathfrak{gl}(4)$ is simply the case of infinitesimal affine transformations. This would relate to Lie equations of the form:

$$\frac{\partial X^\mu}{\partial x^\nu} = a^\mu_\nu, \tag{4.50}$$

for which the matrix on the right-hand side is constant. The solutions then take the form:

$$X^\mu = \varepsilon^\mu + a^\mu_\nu x^\nu. \tag{4.51}$$

## 5 Pre-metric electromagnetism [24-28]

An early observation of Kottler [**24**] and Cartan [**25**] concerning the mathematical structure of Maxwell's equations for electromagnetism was that the only place that the spacetime metric played an essential role was in the definition of what is now called the Hodge star isomorphism, as it is applied to 2-forms. The possibility that one could introduce such an isomorphism independently of the introduction of a spacetime metric was further developed by Van Dantzig [**26**] in a manner that suggested that the fundamental spacetime structure in the eyes of electromagnetism was not a Lorentzian pseudometric on the tangent bundle, but an electromagnetic constitutive law on the bundle of 2-forms.

---

[15] Or simply cf. Guillemin and Sternberg [**20**].

Besides this purely mathematical criticism, there is also a compelling physical reason to formulate the laws of electromagnetism in a metric-free fashion: the Lorentzian structure of spacetime is something that appears as a consequence of the way that electromagnetic waves propagate, and since there are electromagnetic fields that are not wavelike, this suggests that one is prematurely introducing a reduction of the generality of the laws of electromagnetism by assuming the existence of a Lorentzian pseudometric from the outset. A more logically defensible path would be to formulate the general laws without the introduction of a metric and then derive the appearance of the metric as a consequence of the way that electromagnetic waves propagate. The summary of this approach that follows is largely consistent with the work of Hehl and Obukhov [**27**], as well as the author [**28**].

We assume only that the spacetime manifold $M$ is four-dimensional, orientable, and has a unit-volume element $\mathcal{V} \in \Lambda^4(M)$ on $T(M)$, i.e., a globally non-zero 4-form. One then has the isomorphisms of Poincaré duality as a consequence:

$$\#: \Lambda_*(M) \to \Lambda^{4-*}(M), \quad \mathbf{a} \mapsto i_\mathbf{a} \mathcal{V}. \tag{5.1}$$

5.1 Basic equations

The basic (vacuum) equations of pre-metric electromagnetism are:

$$dF = 0, \quad d\#\mathfrak{h} = 0, \quad \mathfrak{h} = \chi(F), \tag{5.2}$$

in which $F \in \Lambda^2(M)$ represents the electromagnetic field strengths, $\mathfrak{h} \in \Lambda_2(M)$ represents the electromagnetic excitations, and $\chi\colon \Lambda^2(M) \to \Lambda_2(M)$ is a fiber-preserving diffeomorphism that covers the identity that one refers to as the *electromagnetic constitutive law* for the theory in question.

5.2 Electromagnetic constitutive laws

Since the electromagnetic constitutive law plays such a fundamental role in pre-metric electromagnetism, one must take pains to formulate that structure in a manner that is sufficiently general to include the physically important cases, without being so general as to suggest no new insight into the nature of the known phenomena.

In this regard, one must perhaps distinguish two principal classes of electromagnetic constitutive laws: the "vacuum" class and the "optical" class. The principal difference is that since the optical class of electromagnetic constitutive laws is concerned with the way that electric and magnetic fields interact with material media, and especially how electromagnetic waves propagate in such media, there are numerous complexities, such as dispersion, that are essential to optical problems, but not necessarily to vacuum problems. The complexity that dispersion introduces mathematically is that of a non-local transformation from field strength 2-forms excitation bivectors, that involves convolving with the past time history of the fields in question. A further complication that material media introduce is that sometimes even a local and linear electromagnetic constitutive law might need to involve complex electric permittivities and magnetic permeabilities, usually in the coupling terms, due to such phenomena as the coupling of the motion of the medium to the propagation of electromagnetic waves in the medium (Faraday effect), or natural optical activity. For more details on the physics of matters, see Post [**29**] or Landau et al. [**30**].

In the present study, we shall restrict the generality by concentrating on essentially vacuum electromagnetic constitutive laws, although we shall leave open the possibility that the law is nonlinear by way of the phase transition of vacuum polarization at high field strengths. Hence, the laws that we shall consider will be possibly nonlinear and asymmetric, but real and non-dispersive. Although the justification is presumably that the vacuum electromagnetic constitutive law, which is usually expressed by the two constants $\varepsilon_0$ and $\mu_0$, is more "fundamental" than the optical ones, nevertheless, since one is dealing with phenomena that occur in a realm that is beyond direct observation, one should always leave open the possibility that the macroscopic phenomena might have microscopic precursors.

Furthermore, we shall restrict the possible nonlinearity somewhat by considering only diffeomorphisms of $\Lambda^2(M)$ that can be represented locally as:

$$\chi(x, F) = \chi(x)(\tfrac{1}{2} F_{\mu\nu} dx^\mu \wedge dx^\nu) = \tfrac{1}{2} \chi^{\mu\nu\rho\sigma}(x^\alpha, F_{\beta\gamma}) F_{\rho\sigma} \partial_\mu \wedge \partial_\nu, \qquad (5.3)$$

so $\chi^{\mu\nu\rho\sigma} \in C^\infty(\Lambda^2(U))$. Hence, although $\chi$ might be *nonlinear* as a map *from* $\Lambda^2(M)$ to $\Lambda^2(M)$, it nevertheless can be represented by a *linear* object *on* $\Lambda^2(M)$, namely, a fourth-rank tensor field. If we take the exterior derivative of the 0-forms $\chi^{\mu\nu\rho\sigma}$ then we get:

$$d\chi^{\mu\nu\rho\sigma} = \chi^{\mu\nu\rho\sigma}{}_{,\alpha} \, dx^\alpha + \chi^{\mu\nu\rho\sigma,\,\alpha\beta} \, dF_{\alpha\beta}, \qquad (5.4)$$

in which the comma indicates partial differentiation with respect to the indicated variable.

This allows us to distinguish four types of constitutive laws, in increasing level of generality:

(*i*) *Uniform linear case:* $\chi$ is a linear isomorphism at every point of *M* and its components are constants:

$$d\chi^{\mu\nu\rho\sigma} = 0. \qquad (5.5)$$

This includes the usual classical assumption that the vacuum of space is characterized by the constants $\varepsilon_0$ and $\mu_0$, which describe its electric permittivity and magnetic permeability.

Of course, constancy of components is a condition that is hardly frame-invariant; indeed, it would seem to define a class of special frames for a given electromagnetic constitutive law (if they exist), much as a rest frame is a special class of frames defined by some physical motions. Indeed, the relationship between rest frames for motions and the character of electromagnetic fields in various frames is rather apparent. Of course, since $c_0 = (\varepsilon_0 \mu_0)^{-1/2}$, the two constants must transform inversely in order to preserve the frame-invariance of $c_0$.

(*ii*) *Non-uniform linear case:* $\chi$ is a linear isomorphism at every point of *M*, but its components are not constants, in general:

$$\chi^{\mu\nu\rho\sigma} = \chi^{\mu\nu\rho\sigma}(x^\lambda), \qquad d\chi^{\mu\nu\rho\sigma} = \chi^{\mu\nu\rho\sigma}{}_{,\alpha}\, dx^\alpha. \tag{5.6}$$

This sort of scenario has a more "linear optical" character than the linear uniform case.

(*iii*) *Uniform nonlinear case:* $\chi$ is not a linear isomorphism, but in such a way that the non-linearity is only a function of field strength:

$$\chi^{\mu\nu\rho\sigma} = \chi^{\mu\nu\rho\sigma}(F), \qquad d\chi^{\mu\nu\rho\sigma} = \chi^{\mu\nu\rho\sigma,\,\alpha\beta}\, dF_{\alpha\beta}. \tag{5.7}$$

This case might possibly serve as an effective model for vacuum polarization, as in the Born-Infeld model. It also represents a common description of nonlinear optical media.

(*iv*) *Non-uniform nonlinear case:* This is the case of full generality for which $\chi$ is not a linear isomorphism, and the non-linearity is a function of both position in *M* and the field strength *F*:

$$\chi^{\mu\nu\rho\sigma} = \chi^{\mu\nu\rho\sigma}(x, F), \qquad d\chi^{\mu\nu\rho\sigma} = \chi^{\mu\nu\rho\sigma}{}_{,\lambda}\, dx^\lambda + \chi^{\mu\nu\rho\sigma,\,\alpha\beta}\, dF_{\alpha\beta}. \tag{5.8}$$

5.3 Symmetries of linear electromagnetic constitutive laws

When $\chi: \Lambda^2(M) \to \Lambda_2(M)$ is linear on the fibers, one can also regard $\chi$ as a fourth-rank tensor field on *M*, and the fiberwise linear maps from $\Lambda^2(M)$ to $\Lambda_2(M)$ are in one-to-one linear correspondence with the sections of $\Lambda_2(M) \otimes \Lambda_2(M)$, which can be represented locally by expressions of the form:

$$\chi = \chi^{\mu\nu\rho\sigma}(\partial_\mu \wedge \partial_\nu) \otimes (\partial_\rho \wedge \partial_\sigma). \tag{5.9}$$

One must have:

$$\chi^{\mu\nu\rho\sigma} = -\chi^{\nu\mu\rho\sigma}, \qquad \chi^{\mu\nu\rho\sigma} = -\chi^{\mu\nu\sigma\rho}, \tag{5.10}$$

since $\chi$ maps antisymmetric tensor fields to anti-symmetric tensor fields.

If **a**, **b** $\in \Lambda_2(M)$ are bivector fields then $\chi$ can also be described as a bilinear functional $\chi(\mathbf{a}, \mathbf{b})$. This then brings up the issue of whether this functional is symmetric or anti-symmetric in its arguments. Following Hehl and Obukhov [**27**], we shall assume that $\chi^{\mu\nu\rho\sigma}$ has the most general symmetries. Hence, we assume that $\chi$ is expressible as a sum:

$$\chi = {}^{(1)}\chi + {}^{(2)}\chi + {}^{(3)}\chi, \tag{5.11}$$

in which ${}^{(1)}\chi$ is called the *principal part* of $\chi$, ${}^{(2)}\chi$ is called the *skewon part*, and ${}^{(3)}\chi$ is called the *axion part*. By definition, one has:

$$\begin{aligned}
{}^{(2)}\chi^{\mu\nu\rho\sigma} &= \tfrac{1}{2}(\chi^{\mu\nu\rho\sigma} - \chi^{\rho\sigma\mu\nu}), & (5.12\mathrm{a}) \\
{}^{(3)}\chi^{\mu\nu\rho\sigma} &= \chi^{[\mu\nu\rho\sigma]} = \lambda\, \varepsilon^{\mu\nu\rho\sigma}, \quad \lambda \in C^\infty(M), & (5.12\mathrm{b}) \\
{}^{(1)}\chi^{\mu\nu\rho\sigma} &= \chi^{\mu\nu\rho\sigma} - {}^{(2)}\chi^{\mu\nu\rho\sigma} - {}^{(3)}\chi^{\mu\nu\rho\sigma}. & (5.12\mathrm{c})
\end{aligned}$$

Hence, this decomposition amounts to the decomposition of the *second*-rank tensor field (on $\Lambda_2(M)$) $\chi$ into its symmetric trace-zero part $^{(1)}\chi^{\mu\nu\rho\sigma}$, its anti-symmetric part $^{(2)}\chi^{\mu\nu\rho\sigma}$, and its trace part $^{(3)}\chi^{\mu\nu\rho\sigma}$, which is an irreducible decomposition of the representation of $GL(6; \mathbb{R})$ in $\Lambda_2(M) \otimes \Lambda_2(M)$.

When one composes the map $\chi$ with the Poincaré duality isomorphism #, one obtains a corresponding decomposition of the isomorphism $\kappa = \# \cdot \chi$:

$$\kappa = {}^{(1)}\kappa + {}^{(2)}\kappa + {}^{(3)}\kappa. \tag{5.13}$$

If one regards each component of $\kappa$ as a linear operator from each fiber of $\Lambda_2(M)$ to itself then since the volume element $\mathcal{V}$ defines a scalar product on $\Lambda_2(M)$ by way of:

$$\langle \mathbf{a}, \mathbf{b} \rangle = \mathcal{V}(\mathbf{a} \wedge \mathbf{b}) \tag{5.14}$$

we can speak of the *adjoint* $\kappa^*$ of the operator $\kappa$ relative to this scalar product, which has the defining property:

$$\langle \kappa^*(\mathbf{a}), \mathbf{b} \rangle = \langle \mathbf{a}, \kappa(\mathbf{b}) \rangle, \tag{5.15}$$

i.e:

$$\kappa^*(\mathbf{a}) \wedge \mathbf{b} = \mathbf{a} \wedge \kappa(\mathbf{b}). \tag{5.16}$$

Having said that, we can characterize the decomposition (5.13) as defining the *self-adjoint traceless* part $^{(1)}\kappa$ of $\kappa$; the *skew-adjoint* part $^{(2)}\kappa$; and the *trace* part $^{(3)}\kappa$, which has the property that:

$$^{(3)}\kappa = \lambda I, \qquad \text{for some } \lambda \in C^\infty(M). \tag{5.17}$$

Of particular interest later are the various traces that one obtains by contraction of the components of $\kappa$. We define:

$$\kappa_\mu{}^\rho \equiv \kappa_{\mu\lambda}{}^{\rho\lambda}, \qquad \kappa \equiv \kappa_\mu{}^\mu, \qquad \mathcal{K}_\mu{}^\rho \equiv \kappa_\mu{}^\rho - \tfrac{1}{4}\kappa \delta_\mu^\rho. \tag{5.18}$$

Since $\kappa_\mu{}^\rho$ and $\mathcal{K}_\mu{}^\rho$ are 4×4 real matrices one can regard them as elements of the Lie algebras $\mathfrak{gl}(4; \mathbb{R})$ and $\mathfrak{sl}(4; \mathbb{R})$, respectively.

There is a subtlety concerned with the isomorphism $\kappa: \Lambda^2(M) \to \Lambda^2(M)$ that must be addressed here, since it affects the matter of the symmetries of the pre-metric equations of electromagnetism: since $\kappa$ includes the Poincaré isomorphism #, which is due to the choice of unit-volume element $\mathcal{V}$, the map $\kappa$ is not actually equivariant under the action of all of $GL(4; \mathbb{R})$, but only the subgroup $SL(4; \mathbb{R})$ that preserves $\mathcal{V}$. This really mostly affects the transformations of the components of 2-forms under changes of local linear frames, in which one says that a general linear transformation $A$ introduces a factor of $\det(A)$ into the transformation law. This property was referred to in the past as saying that a 2-form of the form $\kappa(F)$ is a "tensor density of weight 1," although nowadays other expressions exist, such as calling it a "twisted" 2-form or, in the terminology of Kiehn [**31**], that $F$ is a "pair" 2-form and $k(F)$ is an "impair" 2-form ([16]).

In the conventional Lorentzian case the isomorphism $\chi : \Lambda^2(M) \to \Lambda_2(M)$ is proportional to $\iota_g \wedge \iota_g : \Lambda^2(M) \to \Lambda_2(M)$, $\alpha \wedge \beta \mapsto \iota_g \alpha \wedge \iota_g \beta$, whose components relative to $dx^\mu$ are:

$$(\iota_g \wedge \iota_g)^{\mu\nu\alpha\beta} = g^{\mu\alpha} g^{\nu\beta} - g^{\nu\alpha} g^{\mu\beta}. \tag{5.19}$$

Since:

$$(\iota_g \wedge \iota_g)^{\alpha\beta\mu\nu} = g^{\alpha\mu} g^{\beta\nu} - g^{\beta\mu} g^{\alpha\nu} = (\iota_g \wedge \iota_g)^{\mu\nu\alpha\beta} \tag{5.20}$$

the skewon part $^{(2)}\chi$ of this tensor field vanishes, and since:

$$(\iota_g \wedge \iota_g)^{\mu\nu\alpha\beta} \partial_\mu \wedge \partial_\nu \wedge \partial_\alpha \wedge \partial_\beta = 0, \tag{5.21}$$

due to the fact that $g^{\mu\nu}$ is symmetric in $\mu\nu$ while $\partial_\mu \wedge \partial_\nu \wedge \partial_\alpha \wedge \partial_\beta$ is anti-symmetric in all of its index pairs, one also has $^{(3)}\chi = 0$.

The components of $\mathcal{V}$ can be written as:

$$\frac{1}{4!}\sqrt{-g}\,\varepsilon_{\mu\nu\alpha\beta},$$

which makes the components of $\kappa$ equal to:

$$\kappa_{\mu\lambda}{}^{\alpha\beta} = \frac{1}{4!}\sqrt{-g}\,\varepsilon_{\mu\nu\rho\sigma}\,g^{\rho\alpha}\,g^{\sigma\beta}, \tag{5.22}$$

and:

$$\kappa_\mu{}^\alpha \equiv \kappa_{\mu\lambda}{}^{\alpha\lambda} = \frac{1}{4!}\sqrt{-g}\,\varepsilon_{\mu\lambda\rho\sigma}\,g^{\rho\alpha}\,g^{\sigma\lambda} = 0 \tag{5.23}$$

since $\varepsilon_{\mu\lambda\rho\sigma}$ is anti-symmetric in $\sigma\lambda$, while $g^{\sigma\lambda}$ is symmetric.

The conditions under which the more general linear electromagnetic constitutive law $x$ reduces to $\iota_g \wedge \iota_g$ for some metric $g$ of Lorentzian type are quite important to both physics and mathematics. Clearly, it is necessary that $\chi$ must be symmetric and have vanishing trace part. Furthermore, in order for $\kappa$ to be proportional to * for a Lorentzian metric, $\kappa^2$ must be proportional to $-I$.

The physical considerations that relate to this factorizability of $\chi$ grow out of the Fresnel analysis of the singularities of the wave equation that $\chi$ (or $\kappa$) might define by looking at the factorizability of the Fresnel quartic hypersurface. In this study, we shall mention only that a key physical issue is the non-existence of bi-refringence, in which the speed of propagation of an electromagnetic wave depends upon the direction of its wave vector. For more details, the interested are referred to Hehl and Obukhov [**27**].

---

[16] Although "pair" and "impair" are simply the French words that Cartan used instead of "even" and "odd," respectively, nevertheless, there is a possible source of confusion with the terminology for $k$-forms with $k$ being even or odd, respectively.

Another possible physical issue is based the more "thermodynamical" consideration of irreversibility and whether that macroscopic property of complex systems might have a microscopic precursor at the fundamental level, such as a speed of propagation for electromagnetic waves that is different for opposite timelike directions. (For such matters, see Kiehn [**31**].)

5.4 Representation of the pre-metric Maxwell equations as an exterior differential system on $\Lambda^2(M)$

We can incorporate the electromagnetic constitutive law into the equation for $\mathfrak{h}$ and restate the equations (5.2) as:

$$dF = 0, \quad d\kappa(F) = 0 . \tag{5.24}$$

If we represent the volume element locally by:

$$\mathcal{V} = dx^0 \wedge dx^1 \wedge dx^2 \wedge dx^3 = \frac{1}{4!} \varepsilon_{\mu\nu\rho\sigma} dx^\mu \wedge dx^\nu \wedge dx^\rho \wedge dx^\sigma , \tag{5.25}$$

then we can represent $F$ and $\kappa(F) = \#\chi(F)$ as:

$$F = \frac{1}{2} F_{\mu\nu} dx^\mu \wedge dx^\nu , \qquad \kappa(F) = \frac{1}{2} \frac{1}{4!} \kappa_{\mu\nu}{}^{\alpha\beta} F_{\alpha\beta} dx^\mu \wedge dx^\nu , \tag{5.26}$$

and the system (5.24) as:

$$0 = \Theta^1 \equiv dF_{\mu\nu} \wedge dx^\mu \wedge dx^\nu , \tag{5.27a}$$
$$0 = \Theta^2 \equiv \kappa_{\mu\nu}{}^{\alpha\beta} dF_{\alpha\beta} \wedge dx^\mu \wedge dx^\nu + F_{\alpha\beta} d\kappa_{\mu\nu}{}^{\alpha\beta} \wedge dx^\mu \wedge dx^\nu . \tag{5.27b}$$

Furthermore, it will prove convenient to insert an identity transformation into $\Theta^1$ and expand $d\kappa_{\mu\nu}{}^{\alpha\beta}$ into its components along $dx^\mu$ and $dF_{\alpha\beta}$, which makes (5.27a, b) take the form:

$$0 = \Theta^1 \equiv \delta^\alpha_\mu \delta^\beta_\nu dF_{\alpha\beta} \wedge dx^\mu \wedge dx^\nu , \tag{5.28a}$$
$$0 = \Theta^2 \equiv K_{\mu\nu}{}^{\alpha\beta} dF_{\alpha\beta} \wedge dx^\mu \wedge dx^\nu + F_{\alpha\beta} \kappa_{\mu\nu}{}^{\alpha\beta}{}_{,\lambda} dx^\lambda \wedge dx^\mu \wedge dx^\nu , \tag{5.28b}$$

into which we have introduced the "deformed" constitutive tensor:

$$K_{\mu\nu}{}^{\alpha\beta} \equiv \kappa_{\mu\nu}{}^{\alpha\beta} + F_{\kappa\lambda} \kappa_{\mu\nu}{}^{\alpha\beta, \kappa\lambda} , \tag{5.29}$$

which will clearly play a role only in the nonlinear cases.

We recognize the 3-form $\Theta^1$ as being the canonical 3-form $d\Phi$ on $\Lambda^2(M)$. It is clear that for all four cases of $d\kappa_{\mu\nu}{}^{\alpha\beta}$ – as defined by the four cases of $\chi^{\mu\nu\alpha\beta}$ – we can regard both of the 3-forms $\Theta^i$ as elements of $\Lambda^3(\Lambda^2(M))$. Hence, we can represent the system of equations (5.24) by the following exterior differential system on $\Lambda^2(M)$:

$$\Theta^i = 0, \quad i = 1, 2, \qquad (5.30)$$

or, more precisely, the vanishing of the ideal $I$ in $\Lambda^*(\Lambda^2(M))$ that is generated by $\{\Theta^1, \Theta^2\}$. An arbitrary element of $I$ will then look like $\alpha_i \wedge \Theta^i$, where the $\alpha_i$ are exterior differential forms in $\Lambda^*(\Lambda^2(M))$ that are arbitrary except to the extent that since $\Lambda^2(M)$ is a ten-dimensional manifold and the $\Theta^i$ are 3-forms, any $\alpha_i$ whose degree is greater than seven will automatically drive the exterior product to zero.

Sooner or later, we must address the fact that we have chosen to represent the equations of electromagnetism in terms of the field strength 2-form $F$ and excitation bivector $\mathfrak{h}$, instead of a electromagnetic potential 1-form $\phi$ that makes $F = d\phi$. The main reasons are:

*i*) Unless one assumes that spacetime $M$ has vanishing de Rham cohomology in dimension two, $\phi$ will be a local object, not a global one, unless one represents it as a global object with singularities. Hence, one would not really be dealing with the vector bundle of 1-forms – i.e., the cotangent bundle – on spacetime, but the sheaf of germs of local 1-forms, which would make defining the differential system more involved.

*ii*) The electromagnetic constitutive law is more fundamentally related to the field strengths and excitations than to a choice of potential 1-form.

*iii*) In mechanics, one generally regards forces as more fundamental than potentials, or, more precisely, potential *differences*. Originally, electric and magnetic field strengths are defined a forces, whereas the electric and magnetic potentials are defined in terms of potential energy.

Nevertheless, if one chooses to leave out the role of electromagnetic potentials then one must then accept the consequence that one is also omitting the symmetry of the system that takes the form of gauge symmetry. However, in many treatments of the symmetries of electromagnetism, the gauge symmetry is regarded as, in a sense, "non-local" as compared to the symmetries that act on spacetime or the field space directly.

### 5.5 Integrability of the system

Since both of the generating 3-forms $\Theta^1$ and $\Theta^2$ of our exterior differential system are closed, the only possible obstruction to the integrability of the system (5.28a, b) would be the possibility that the integral elements of the system do not have constant dimension at every $F \in \Lambda^2(M)$. Hence, we should look at the dimensions of the annihilating subspaces to $\Theta^1$ and $\Theta^2$, and since an integral element is the intersection of the two annihilating subspaces, the dimension of their intersection. Since a solution to the system is a submanifold $F: M \to \Lambda^2(M)$ whose induced tangent spaces in $T(\Lambda^2(M))$ must be contained in these intersections and have dimension four, we immediately see that the dimension of the intersections must be at least four.

Now, since the manifold $\Lambda^2(M)$ has dimension ten, the dimension of the annihilating subspace of a non-zero 3-form on $\Lambda^2(M)$ will be seven. The dimension of the join of the two spaces cannot exceed ten, so the dimension of the intersection cannot be less than four. Hence, naively, the dimension of an integral element of our system can range from four to seven. Since one of the 3-forms, namely $\Theta^1 = d\Phi$ is canonically defined on $\Lambda^2(M)$ the only possible source of variance in this dimension will be the character of the constitutive law $\kappa$ at each point.

The case of and intersection of dimension four is attained when the annihilating subspaces of $\Theta^1$ and $\Theta^2$ are transverse; the opposite extreme of dimension seven is attained when they are identical. When the annihilating subspaces of $\Theta^1$ and $\Theta^2$ are transverse they collectively define a tangent subspace that is the annihilating subspace of the 6-form $\Theta^1 \wedge \Theta^2$, which must then be non-vanishing. Now:

$$\begin{aligned}\Theta^1 \wedge \Theta^2 &= K_{\mu\nu}{}^{\alpha\beta}\, dF_{\alpha\beta} \wedge dF_{\gamma\delta} \wedge dx^\mu \wedge dx^\nu \wedge dx^\gamma \wedge dx^\delta \\ &\quad + F_{\alpha\beta}\, \kappa_{\gamma\delta}{}^{\alpha\beta}{}_{,\lambda}\, dF_{\mu\nu} \wedge dx^\lambda \wedge dx^\mu \wedge dx^\nu \wedge dx^\gamma \wedge dx^\delta . \\ &= K_{\mu\nu}{}^{\alpha\beta}\, dF_{\alpha\beta} \wedge dF_{\gamma\delta} \wedge dx^\mu \wedge dx^\nu \wedge dx^\gamma \wedge dx^\delta .\end{aligned} \qquad (5.31)$$

The last step derives from the fact that there are no non-zero 5-forms on a four-dimensional manifold.

For linear $\kappa$'s – for which, $d\kappa_{\mu\nu}{}^{\alpha\beta} = \kappa_{\mu\nu}{}^{\alpha\beta}{}_{,\kappa}\, dx^\kappa$ – one will have:

$$\Theta^1 \wedge \Theta^2 = \kappa_{\mu\nu}{}^{\alpha\beta}\, dF_{\alpha\beta} \wedge dF_{\gamma\delta} \wedge dx^\mu \wedge dx^\nu \wedge dx^\gamma \wedge dx^\delta , \qquad (5.32)$$

which is non-vanishing since $\kappa$ is an isomorphism and all of the factors in the exterior product are linearly independent for suitable choices of the indices. Hence, $\Theta^1 \wedge \Theta^2$ has maximal rank at every point, and the system (5.28a,b) is completely integrable for any choice of $\kappa$ when $\kappa$ is linear.

The term involving $K_{\mu\nu}{}^{\alpha\beta}$ will differ $\kappa_{\mu\nu}{}^{\alpha\beta}$ from only if $d\kappa_{\gamma\delta}{}^{\alpha\beta}$ has a non-vanishing contribution from the $dF_{\mu\nu}$; i.e., $\chi = \chi(F)$ or $\chi = \chi(x, F)$, which is the case of a nonlinear $\kappa$. The expression $\Theta^1 \wedge \Theta^2$ will vanish for $\kappa$'s that satisfy:

$$K_{\mu\nu}{}^{\alpha\beta} = 0 , \qquad (5.33)$$

which suggests that, depending upon the character of $\kappa_{\mu\nu}{}^{\alpha\beta,\kappa\lambda}$, there might be regimes of $F_{\kappa\lambda}$ for which the deformation $F_{\kappa\lambda}\, \kappa_{\mu\nu}{}^{\alpha\beta,\kappa\lambda}$ destroys the invertibility of the map $\kappa$. Hence, singularities in the dimension of the annihilating subspaces for $\Theta^1 \wedge \Theta^2$ might possibly obstruct the global integrability of the nonlinear pre-metric Maxwell equations depending upon the specific nature of $\kappa$; in particular, this would affect $\kappa$'s that depend upon the $F_{\kappa\lambda}$.

### 5.6 General symmetries of the pre-metric Maxwell system

An infinitesimal symmetry of the exterior differential system (5.28a, b) is a vector field $X$ on $\Lambda^2(M)$ with the property that $L_X I \subset I$. Hence, since $\Theta^i$ are both closed, $X$ must satisfy:

$$L_X \Theta^i = d i_X \Theta^i = \alpha_i\, \Theta^1 + \beta_i\, \Theta^2 \qquad (5.34)$$

for some appropriate $\alpha_i$, $\beta_i \in C^\infty(\Lambda^2(M))$.

Since both of the Lie derivatives are 3-forms, which are also the forms of least degree in the generating set, the right-hand side of (5.34) takes the form of the typical element of the $C^\infty(\Lambda^2(M))$-module $\mathcal{M}$ that is spanned by $\Theta^1$ and $\Theta^2$, which will take the form:

$$\alpha\, dF_{\mu\nu} \wedge dx^\mu \wedge dx^\nu + \beta\,(\kappa_{\mu\nu}{}^{\alpha\beta}\, dF_{\alpha\beta} \wedge dx^\mu \wedge dx^\nu + F_{\alpha\beta}\, d\kappa_{\mu\nu}{}^{\alpha\beta} \wedge dx^\mu \wedge dx^\nu)$$
$$= [\alpha \delta^\alpha{}_\mu \delta^\beta{}_\nu + \beta\,(\kappa_{\mu\nu}{}^{\alpha\beta} + F_{\kappa\lambda}\,\kappa_{\mu\nu}{}^{\kappa\lambda,\,\alpha\beta})]\, dF_{\alpha\beta} \wedge dx^\mu \wedge dx^\nu$$
$$+ \beta\, F_{\alpha\beta}\, \kappa_{\mu\nu}{}^{\alpha\beta}{}_{,\lambda}\, dx^\lambda \wedge dx^\mu \wedge dx^\nu. \tag{5.35}$$

Here, we shall introduce two notations: We rewrite (5.34) in the form:

$$L_X \Theta^i \equiv 0 \pmod{\mathcal{M}} \tag{5.36}$$

and define the rank-two $C^\infty(\Lambda^2(M))$-module $\mathcal{C}$, which consists of all expressions of the form $\alpha\, \delta^\alpha{}_\mu \delta^\beta{}_\nu + \beta\, \kappa_{\mu\nu}{}^{\alpha\beta}$, with $\alpha$, $\beta$ being smooth functions on $\Lambda^2(M)$. Our choice of notation is suggestive of the fact that since the constitutive laws that we shall be interested in are conformal to almost complex structures on $\Lambda^2(M)$, the elements of $\mathcal{C}$, which take the component-free form $\alpha I + \beta \kappa$, can be associated with complex-valued functions on $\Lambda^2(M)$ by the $\mathbb{C}$-linear isomorphism that takes $\alpha I + \beta \kappa$ to $\alpha + i\beta$.

If we express $X$ in local form as:

$$X = X^\mu \frac{\partial}{\partial x^\mu} + X_{\mu\nu} \frac{\partial}{\partial F_{\mu\nu}} \tag{5.37}$$

and use the local form of the $\Theta^i$ that is given by (5.28a, b) then the infinitesimal symmetries are defined by the pair of equations:

$$L_X \Theta^1 = L_X (\delta^\alpha{}_\mu \delta^\beta{}_\nu\, dF_{\alpha\beta} \wedge dx^\mu \wedge dx^\nu) = \delta^\alpha{}_\mu \delta^\beta{}_\nu\, L_X (dF_{\alpha\beta} \wedge dx^\mu \wedge dx^\nu)$$
$$= X_{\mu\nu,\lambda}\, dx^\lambda \wedge dx^\mu \wedge dx^\nu + [2X^\alpha{}_{,\lambda} \delta^\beta{}_\nu + X_{\mu\nu}{}^{,\alpha\beta}]\, dF_{\alpha\beta} \wedge dx^\mu \wedge dx^\nu$$
$$+ X^{\mu,\alpha\beta}\, dF_{\mu\nu} \wedge dF_{\alpha\beta} \wedge dx^\nu$$
$$\equiv 0 \pmod{\mathcal{M}} \tag{5.38a}$$

$$L_X \Theta^2 = (L_X \kappa_{\mu\nu}{}^{\alpha\beta})\, dF_{\alpha\beta} \wedge dx^\mu \wedge dx^\nu + \kappa_{\mu\nu}{}^{\alpha\beta}\, L_X (dF_{\alpha\beta} \wedge dx^\mu \wedge dx^\nu)$$
$$+ L_X (F_{\alpha\beta}\, \kappa_{\mu\nu}{}^{\alpha\beta}{}_{,\lambda}\, dx^\lambda \wedge dx^\mu \wedge dx^\nu)$$

$$= (L_X \kappa_{\mu\nu}{}^{\alpha\beta})\, dF_{\alpha\beta} \wedge dx^\mu \wedge dx^\nu + \kappa_{\mu\nu}{}^{\alpha\beta}\, L_X (dF_{\alpha\beta} \wedge dx^\mu \wedge dx^\nu)$$
$$+ X_{\alpha\beta}\, \kappa_{\mu\nu}{}^{\alpha\beta}{}_{,\lambda}\, dx^\lambda \wedge dx^\mu \wedge dx^\nu + (L_X \kappa_{\mu\nu}{}^{\alpha\beta}{}_{,\lambda})\, F_{\alpha\beta}\, dx^\lambda \wedge dx^\mu \wedge dx^\nu$$
$$+ F_{\alpha\beta}\, \kappa_{\mu\nu}{}^{\alpha\beta}{}_{,\lambda}\, L_X (dx^\lambda \wedge dx^\mu \wedge dx^\nu)$$
$$\equiv 0 \pmod{\mathcal{M}} \tag{5.38b}$$

From the first equation, we derive the system:

$$X_{\mu\nu,\lambda} = \beta\, F_{\alpha\beta}\, \kappa_{\mu\nu}{}^{\alpha\beta}{}_{,\lambda}, \tag{5.39a}$$
$$2X^\alpha{}_{,\lambda} \delta^\beta{}_\nu + X_{\mu\nu}{}^{,\alpha\beta} = \alpha \delta^\alpha{}_\mu \delta^\beta{}_\nu + \beta\, \kappa_{\mu\nu}{}^{\alpha\beta}, \tag{5.39b}$$
$$X^{\mu,\alpha\beta} = 0. \tag{5.39c}$$

The last of these three says that $X^\mu = X^\mu(x^\mu)$ and simplifies the computations in $L_X \Theta^2$ by eliminating all of the terms that involve the affected derivatives.

Let us examine the individual terms of $L_X \Theta^2$. The first term gives us that in order for the resulting over-determined system of partial differential equations for $X$ to be consistent, we must introduce the compatibility condition:

$$L_X K_{\mu\nu}{}^{\alpha\beta} \equiv 0 \pmod{\mathcal{C}}, \tag{5.40}$$

i.e.:

$$L_X K_{\mu\nu}{}^{\alpha\beta} = \alpha\, \delta^\alpha_\mu\, \delta^\beta_\nu + \beta\, K_{\mu\nu}{}^{\alpha\beta}. \tag{5.41}$$

This equation relates to a sort of complex generalization of the conformal Killing equation for the metric on a Riemannian or Lorentzian manifold. We shall return to a discussion of this equation at the end of the section.

The second term satisfies:

$$K_{\mu\nu}{}^{\alpha\beta} L_X (dF_{\alpha\beta} \wedge dx^\mu \wedge dx^\nu) \equiv 0 \pmod{\mathcal{M}}, \tag{5.42}$$

at least when $K_{\mu\nu}{}^{\alpha\beta}$ is invertible, since $\mathcal{M}$ is closed under the action of the isomorphism $K$ and the fact that $L_X (dF_{\alpha\beta} \wedge dx^\mu \wedge dx^\nu) \equiv 0 \pmod{\mathcal{M}}$ follows from (5.38a).

The third term gives the condition that:

$$X_{\alpha\beta}\, \kappa_{\mu\nu}{}^{\alpha\beta}{}_{,\lambda} = \gamma_1 F_{\alpha\beta}\, \kappa_{\mu\nu}{}^{\alpha\beta}{}_{,\lambda}, \tag{5.43}$$

for some smooth function $\gamma_1$ on $\Lambda^2(\mathbb{R}^4)$. Note that this constraint affects the components of $X$ only for $\kappa$'s that depend upon $dx^\mu$.

The fourth term suggests a further restriction on the Lie derivative of $\kappa$; namely a restriction on the Lie derivative of its first partial derivative with respect to $x^\mu$:

$$L_X \kappa_{\mu\nu}{}^{\alpha\beta}{}_{,\lambda} = \gamma_2\, \kappa_{\mu\nu}{}^{\alpha\beta}{}_{,\lambda}, \tag{5.44}$$

for some other smooth function $\gamma_2$ on $\Lambda^2(\mathbb{R}^4)$.

The final term gives:

$$F_{\alpha\beta}\, \kappa_{\mu\nu}{}^{\alpha\beta}{}_{,\lambda}\, L_X (dx^\lambda \wedge dx^\mu \wedge dx^\nu) = \gamma_3\, F_{\alpha\beta}\, \kappa_{\mu\nu}{}^{\alpha\beta}{}_{,\lambda}\, dx^\lambda \wedge dx^\mu \wedge dx^\nu, \tag{5.45}$$

for some a smooth function $\gamma_3$ on $\Lambda^2(\mathbb{R}^4)$. Now:

$$\begin{aligned} L_X (dx^\lambda \wedge dx^\mu \wedge dx^\nu) &= dX^\lambda \wedge dx^\mu \wedge dx^\nu + dx^\lambda \wedge dX^\mu \wedge dx^\nu + dx^\lambda \wedge dx^\mu \wedge dX^\nu \\ &= (X^\lambda{}_{,\alpha}\, \delta^\mu_\beta\, \delta^\nu_\gamma + \delta^\lambda_\alpha\, X^\mu{}_{,\beta}\, \delta^\nu_\gamma + \delta^\lambda_\alpha\, \delta^\mu_\beta X^\nu{}_{,\gamma})\, dx^\alpha \wedge dx^\beta \wedge dx^\gamma, \end{aligned} \tag{5.46}$$

so if we are to satisfy (5.44), we must have:

$$F_{\rho\sigma}(X^\lambda{}_{,\alpha}\, \kappa_{\beta\gamma}{}^{\rho\sigma}{}_{,\lambda} + X^\lambda{}_{,\beta}\, \kappa_{\lambda\gamma}{}^{\rho\sigma}{}_{,\alpha} + X^\lambda{}_{,\gamma}\, \kappa_{\beta\lambda}{}^{\rho\sigma}{}_{,\alpha}) = \gamma_3\, F_{\rho\sigma}\, \kappa_{\alpha\beta}{}^{\rho\sigma}{}_{,\gamma}, \tag{5.47}$$

for some $\gamma_3$.

We should summarize the infinitesimal symmetry equations thus far. However, we will distinguish the three cases of $d\kappa_{\mu\nu}{}^{\alpha\beta}$. When $d\kappa_{\mu\nu}{}^{\alpha\beta} = 0$, we have:

$$2X^{\alpha}{}_{,\lambda}\,\delta^{\beta}{}_{\nu} + X_{\mu\nu}{}^{,\alpha\beta} = \alpha\delta^{\alpha}{}_{\mu}\,\delta^{\beta}{}_{\nu} + \beta\,\kappa_{\mu\nu}{}^{\alpha\beta}, \tag{5.48a}$$
$$X^{\mu,\,\alpha\beta} = 0. \tag{5.48b}$$
$$L_X \kappa_{\mu\nu}{}^{\alpha\beta} \equiv 0 \;(\text{mod } \mathscr{C}), \tag{5.48c}$$

When $\kappa_{\mu\nu}{}^{\alpha\beta}{}_{,\lambda} \neq 0$, and $\kappa_{\mu\nu}{}^{\alpha\beta,\kappa\lambda} = 0$ we have:

$$2X^{\alpha}{}_{,\lambda}\,\delta^{\beta}{}_{\nu} + X_{\mu\nu}{}^{,\alpha\beta} = \alpha\delta^{\alpha}{}_{\mu}\,\delta^{\beta}{}_{\nu} + \beta\,\kappa_{\mu\nu}{}^{\alpha\beta}, \tag{5.48a}$$
$$X^{\mu,\,\alpha\beta} = 0. \tag{5.48b}$$
$$L_X \kappa_{\mu\nu}{}^{\alpha\beta} \equiv 0 \;(\text{mod } \mathscr{C}), \tag{5.49c}$$
$$L_X \kappa_{\mu\nu}{}^{\alpha\beta}{}_{,\lambda} = \gamma_2 \kappa_{\mu\nu}{}^{\alpha\beta}{}_{,\lambda}, \tag{5.49d}$$
$$X_{\mu\nu,\lambda} = \beta\, F_{\alpha\beta}\, \kappa_{\mu\nu}{}^{\alpha\beta}{}_{,\lambda}, \tag{5.49e}$$
$$X_{\alpha\beta}\, \kappa_{\mu\nu}{}^{\alpha\beta}{}_{,\lambda} = \beta\, F_{\alpha\beta}\, \kappa_{\mu\nu}{}^{\alpha\beta}{}_{,\lambda}, \tag{5.49f}$$
$$F_{\rho\sigma}(X^{\lambda}{}_{,\alpha}\, \kappa_{\beta\gamma}{}^{\rho\sigma}{}_{,\lambda} + X^{\mu}{}_{,\beta}\, \kappa_{\mu\gamma}{}^{\rho\sigma}{}_{,\alpha} + X^{\nu}{}_{,\gamma}\, \kappa_{\beta\nu}{}^{\rho\sigma}{}_{,\alpha}) = \gamma_3\, F_{\rho\sigma}\, \kappa_{\alpha\beta}{}^{\rho\sigma}{}_{,\gamma}. \tag{5.49g}$$

When $\kappa_{\mu\nu}{}^{\alpha\beta}{}_{,\lambda} = 0$, but, $\kappa_{\mu\nu}{}^{\alpha\beta,\kappa\lambda} \neq 0$, we have:,

$$2X^{\alpha}{}_{,\lambda}\,\delta^{\beta}{}_{\nu} + X_{\mu\nu}{}^{,\alpha\beta} = \alpha\delta^{\alpha}{}_{\mu}\,\delta^{\beta}{}_{\nu} + \beta\, K_{\mu\nu}{}^{\alpha\beta}, \tag{5.50a}$$
$$X^{\mu,\,\alpha\beta} = 0. \tag{5.50b}$$
$$L_X K_{\mu\nu}{}^{\alpha\beta} \equiv 0 \;(\text{mod } \mathscr{C}), \tag{5.50c}$$

If we recall that $\kappa = \# \cdot \chi$ – i.e., $\kappa_{\mu\nu}{}^{\alpha\beta} = \varepsilon_{\mu\nu\kappa\lambda}\,\chi^{\kappa\lambda\alpha\beta}$ – then we see that when $\kappa_{\mu\nu}{}^{\kappa\lambda,\,\alpha\beta} = 0$, (5.40) expands to:

$$(L_X \varepsilon_{\mu\nu\kappa\lambda})\chi^{\kappa\lambda\alpha\beta} + \varepsilon_{\mu\nu\kappa\lambda}(L_X\chi^{\kappa\lambda\alpha\beta}) = \alpha_3\, \delta^{\alpha}{}_{\mu}\,\delta^{\beta}{}_{\nu} + \beta_3\, \varepsilon_{\mu\nu\kappa\lambda}\,\chi^{\kappa\lambda\alpha\beta}. \tag{5.51}$$

This splits into two equations:

$$L_X \varepsilon_{\mu\nu\kappa\lambda} = \alpha_4 \chi_{\mu\nu\kappa\lambda} + \beta_4 \varepsilon_{\mu\nu\kappa\lambda}, \tag{5.52a}$$
$$L_X \chi^{\kappa\lambda\alpha\beta} = \alpha_5\, \varepsilon^{\kappa\lambda\alpha\beta} + \beta_5 \chi^{\kappa\lambda\alpha\beta}, \tag{5.52b}$$

for suitable functions $\alpha_4$, $\alpha_5$, $\beta_4$, $\beta_5$ on $\Lambda^2(M)$. The first one is satisfied by any vector field $X$; one simply uses $\alpha_4 = 0$, $\beta_4 = \partial X^{\mu}/\partial x^{\mu}$.

The second one of the pair is related to a complex generalization of the conformal Killing equation for $\chi$, which can, as we pointed out, define a fiber metric on $\Lambda^2(M)$ when it is symmetric. Hence, of the two components to $\kappa$, namely, # and $\chi$, it is $\chi$ that affects the symmetries of the pre-metric Maxwell system.

One needs to distinguish between the expression $L_X\chi^{\kappa\lambda\alpha\beta}$, which equals $i_X d\chi^{\kappa\lambda\alpha\beta}$, and the expression $(L_X\chi)^{\kappa\lambda\alpha\beta}$, which represents the components of $L_X\chi$ for the given choice of frame. In order to evaluate the local form for $(L_X\chi)^{\kappa\lambda\alpha\beta}$, we form the tensor field:

$$\chi = \tfrac{1}{4}\chi^{\kappa\lambda\alpha\beta}\,\partial_\kappa \wedge \partial_\lambda \otimes \partial_\alpha \wedge \partial_\beta, \tag{5.53}$$

and take the Lie derivative along $X$:

$$L_X \chi = \tfrac{1}{4} \{i_X d\chi^{\kappa\lambda\alpha\beta} \\ - X^{\kappa}{}_{,\rho}\chi^{\rho\lambda\alpha\beta} - X^{\lambda}{}_{,\rho}\chi^{\kappa\rho\alpha\beta} - X^{\alpha}{}_{,\rho}\chi^{\kappa\lambda\rho\beta} - X^{\beta}{}_{,\rho}\chi^{\kappa\lambda\alpha\rho}\}\partial_\kappa \wedge \partial_\lambda \otimes \partial_\alpha \wedge \partial_\beta. \qquad (5.54)$$

Hence:

$$L_X \chi^{\kappa\lambda\alpha\beta} = i_X d\chi^{\kappa\lambda\alpha\beta} - X^{\kappa}{}_{,\rho}\chi^{\rho\lambda\alpha\beta} - X^{\lambda}{}_{,\rho}\chi^{\kappa\rho\alpha\beta} - X^{\alpha}{}_{,\rho}\chi^{\kappa\lambda\rho\beta} - X^{\beta}{}_{,\rho}\chi^{\kappa\lambda\alpha\rho}. \qquad (5.55)$$

In particular, note that this does not have to vanish when the components $\chi^{\kappa\lambda\alpha\beta}$ are constant.

Now, we shall examine the form that aforementioned infinitesimal symmetry equations take by choosing some simple forms for $d\kappa_{\mu\nu}{}^{\alpha\beta}$ that are of immediate physical significance.

## 6 Symmetries for particular electromagnetic constitutive laws

We now examine the consequences of assuming each successive type of electromagnetic constitutive law that was defined above, except for case (*iv*), which represented full generality.

### 6.1 Uniform linear case

For this type of constitutive law:

$$d\kappa_{\mu\nu}{}^{\alpha\beta} = 0, \qquad (6.1)$$

and the pre-metric Maxwell equations become:

$$0 = \Theta^1 = dF_{\mu\nu} \wedge dx^\mu \wedge dx^\nu, \qquad (6.2a)$$
$$0 = \Theta^2 = \kappa_{\mu\nu}^{\alpha\beta} dF_{\alpha\beta} \wedge dx^\mu \wedge dx^\nu. \qquad (6.2b)$$

Hence, the $C^\infty(\Lambda^2(M))$-module $\mathcal{M}$ generated by $\{\Theta^1, \Theta^2\}$ consists of all 3-forms of the form:

$$\alpha\Theta^1 + \beta\Theta^2 = (\alpha\, \delta^\alpha{}_\mu \delta^\beta{}_\nu + \beta\, \kappa_{\mu\nu}{}^{\alpha\beta})\, dF_{\alpha\beta} \wedge dx^\mu \wedge dx^\nu. \qquad (6.3)$$

with $\alpha, \beta \in C^\infty(\Lambda^2(M))$.

The infinitesimal symmetry equations from that are appropriate for this type of $\kappa$ are derived from (5.48a-c):

$$2\frac{\partial X^\alpha}{\partial x^\mu} \delta^\beta{}_\nu = \alpha_1 \delta^\alpha{}_\mu \delta^\beta{}_\nu + \beta_1\, \kappa_{\mu\nu}{}^{\alpha\beta}, \qquad (6.4a)$$

$$\frac{\partial X_{\mu\nu}}{\partial F_{\alpha\beta}} = \alpha_2\, \delta^\alpha{}_\mu \delta^\beta{}_\nu + \beta_2\, \kappa_{\mu\nu}{}^{\alpha\beta}, \qquad (6.4b)$$

$$L_X \kappa_{\mu\nu}{}^{\alpha\beta} = \alpha_3\, \delta^\alpha{}_\mu \delta^\beta{}_\nu + \beta_3\, \kappa_{\mu\nu}{}^{\alpha\beta}, \qquad (6.4c)$$

$$X^\mu = X^\mu(x^\lambda), \qquad X_{\mu\nu} = X_{\mu\nu}(F_{\mu\nu}). \tag{6.4d}$$

As long as the matrix $\alpha^\beta{}_\nu(x^\lambda) \in C^\infty(\Lambda^2(\mathbb{R}^4))$ has non-zero trace, we can right-multiply both sides of (6.4a) by it to obtain the system:

$$\frac{\partial X^\alpha}{\partial x^\mu} = \alpha\,\delta^\alpha{}_\mu + \beta^\alpha{}_\mu \tag{6.5}$$

into which we have introduced:

$$\alpha \equiv \alpha^\beta{}_\beta, \qquad \beta^\alpha{}_\mu \equiv \kappa_{\mu\beta}{}^{\alpha\nu}\alpha^\beta{}_\nu. \tag{6.6}$$

Since the functions $\alpha^\beta{}_\nu$ are arbitrary, and the right-hand side of (6.5) takes the form of the decomposition of a matrix in $\mathfrak{gl}(4;\mathbb{R})$ into its trace part and trace-free part in $\mathfrak{sl}(4;\mathbb{R})$, we can also assume that $\alpha^\alpha{}_\mu \in \mathfrak{sl}(4;\mathbb{R})$. Collectively, the right-hand side of (6.5) defines an element of the Lie algebra of $GL(4;\mathbb{R})$ for each $(x^\lambda, F_{\mu\nu}) \in \Lambda^2(\mathbb{R}^4)$.

From what we said in section 4.2, in order to integrate (6.5) and obtain the infinitesimal generator of a one-parameter subgroup of Diff($M$) that represents some element of $\mathfrak{gl}(4;\mathbb{R})$ we see that we are actually confronting four possibilities, depending upon the nature of the functions $\alpha$ and $\beta^\alpha{}_\mu$, since we will be looking at all of the possible ways of extending the formal algebra $\mathbb{R}^4 \oplus \mathfrak{gl}(4;\mathbb{R})$. From simplest to most complicated, one has the affine Lie algebra $\mathbb{R}^4 \oplus \mathfrak{gl}(4;\mathbb{R})$, the formal algebra $\mathbb{R}^4 \oplus \mathfrak{gl}(4) \oplus \mathfrak{p}^{(1)}$, which also represents the Lie algebra $\mathfrak{sl}(5;\mathbb{R})$, the formal algebra $\{\mathfrak{gl}(4), \mathfrak{sl}(4)^{(1)}\}[[4]]$, which gives vector fields of constant divergence, and the entire formal algebra $\mathfrak{gl}[[4]]$.

In the first case, $\alpha$ and $\beta^\alpha{}_\mu$ are constant as functions of $x^\lambda$, and one can integrate (6.5) immediately and obtain:

$$X^\alpha = \varepsilon^\alpha + \alpha x^\alpha + \alpha^\alpha{}_\mu x^\mu, \tag{6.7}$$

in which the $\varepsilon^\mu$ are also constant as functions of $x^\lambda$. Hence, in this case we have deduced that the spacetime dilatations and translations are symmetries of the system (6.2a,b), as well as the special linear transformations whose infinitesimal generator is $\alpha^\alpha{}_\mu$.

The second possibility, that we are generally dealing with the formal algebra $\mathbb{R}^4 \oplus \mathfrak{gl}(4) \oplus \mathfrak{p}^{(1)}$, means that we can only allow $\alpha$ and $\beta^\alpha{}_\mu$ that are linear in $x^\lambda$, which adds the four extra generators that originate in $\mathfrak{p}^{(1)}$:

$$X^\mu = (\alpha_\nu x^\nu)x^\mu, \tag{6.8}$$

to the generators described by (6.7). The four generators described by (6.8) represent infinitesimal inversions through a hyperplane defined by $\alpha_\nu$.

The other two possibilities, being of infinite order, are harder to describe in such concise terms, except to say that one is either concerned with all vector fields on $\mathbb{R}^4$ of constant divergence or all vector fields, in general.

From a physical standpoint the second possibility seems the most intriguing since it seems to extend the scope of the known symmetries of electromagnetism, which we shall demonstrate shortly, in a natural way without producing an overly general sort of consequence. It suggests that perhaps the most promising direction for the further geometrization of electromagnetism is projective differential geometry.

Now, let us consider the second set of Lie equations (6.4b), which concern the fiber transformations. When $\alpha_2$, $\beta_2$ are constant as functions of $F_{\mu\nu}$, equation (6.4b) integrates to:

$$X_{\mu\nu} = \phi_{\mu\nu} + \alpha_2 F_{\mu\nu} + \beta_2 \kappa_{\mu\nu}{}^{\alpha\beta} F_{\alpha\beta}, \tag{6.9}$$

in which the $\phi_{\mu\nu}$ are constant. Hence, we have deduced that the field-space (i.e., fiber) translations, field-space dilatations, and the "duality transformation" defined by the matrix $\kappa_{\mu\nu}{}^{\alpha\beta}$ are symmetries of our system, as well. Once again, since our transformations can only take solutions to solutions, the dilatations and translations are a predictable artifact of the linearity of the pre-metric field equations in the present case, since it corresponds to the addition of any other solution to a given solution.

Actually, since the isomorphism $\kappa$ is proportional to an almost-complex structure on $\Lambda^2(\mathbb{R}^4)$, we can re-express the right-hand side of (6.4b) in the form:

$$z = \alpha_2 I + \beta_2 *, \tag{6.10}$$

in which the $I$ in question is the identity transformation for 2-forms. Hence, the dilatations and duality rotations can be combined into a representation of the multiplicative group $\mathbb{C}^*$ in $GL(\Lambda^2(\mathbb{R}^4))$. Its Lie algebra is $(\mathbb{C}, +)$, which can also be regarded in real form as $\mathbb{R} \oplus \mathfrak{so}(2)$.

It is clear that we cannot find solutions to (6.4b) for which the functions $\alpha_2$, $\beta_2$ are non-constant functions of $F_{\mu\nu}$, since $\kappa_{\mu\nu}{}^{\alpha\beta}$ is not such a function. For instance, if we try:

$$X_{\mu\nu} = G^{\kappa\lambda} F_{\kappa\lambda} F_{\mu\nu}, \tag{6.11}$$

for some constants $G^{\alpha\beta}$ so:

$$\frac{\partial X_{\mu\nu}}{\partial F_{\alpha\beta}} = G^{\alpha\beta} F_{\mu\nu} + (G^{\kappa\lambda} F_{\kappa\lambda}) \delta^{\alpha}{}_{\mu} \delta^{\beta}{}_{\nu}, \tag{6.12}$$

then the only way that this equals $\alpha \delta^{\alpha}{}_{\mu} \delta^{\beta}{}_{\nu} + \beta \kappa_{\mu\nu}{}^{\alpha\beta}$ for some $a, b$ is if:

$$\kappa_{\mu\nu}{}^{\alpha\beta} = G^{\alpha\beta} F_{\mu\nu}; \tag{6.13}$$

but this would make $\kappa_{\mu\nu}{}^{\alpha\beta}$ a function of $F_{\mu\nu}$.

Now, let us see if we can make contact with known results (cf., Harrison and Estabrook [**3**]) about the symmetries of the linear Maxwell system. We then suppose that the isomorphism $\kappa$ reduces to the Hodge * isomorphism that one obtains from a Lorentzian metric $g$ on $T(M)$ that takes the form:

$$g = \eta_{\mu\nu} dx^{\mu} dx^{\nu} \tag{6.14}$$

in the natural coframe $dx^{\mu}$, which is then assumed to be orthogonal ([17]).

---

[17] We shall pass over the detail that this is only possible if the reduction of the bundle of linear frames on $M$ that $g$ defines is an integrable $SO(3, 1)$-structure, which can be obstructed by topology, just as we passed over the topological obstructions to that reduction itself.

Hence:

$$\kappa_{\mu\nu}{}^{\alpha\beta} = \frac{1}{4!} \varepsilon_{\mu\nu\rho\sigma} \eta^{\rho\alpha} \eta^{\sigma\beta}. \tag{6.15}$$

and:

$$\kappa_{\mu\nu}{}^{\alpha\beta} \beta^{\nu}{}_{\beta} = \frac{1}{4!} \varepsilon_{\mu\nu\rho\sigma} \eta^{\rho\alpha} \beta^{\nu\sigma}. \tag{6.16}$$

When $\beta^{\nu\sigma}$ is symmetric, we get:

$$\beta^{\nu}{}_{\beta} \kappa_{\mu\nu}{}^{\alpha\beta} = 0. \tag{6.17}$$

Hence, only an anti-symmetric set of $\beta^{\nu\sigma}$ will give a non-zero contribution. We set:

$$\alpha_{\mu\nu} = \frac{1}{4!} \varepsilon_{\mu\nu\rho\sigma} \beta^{\rho\sigma}, \tag{6.18}$$

so $\alpha^{\alpha}{}_{\mu} = \eta^{\alpha\lambda} \alpha_{\lambda\mu}$. Hence, since $\alpha_{\mu\nu}$ is anti-symmetric, we can also regard it as an arbitrary infinitesimal four-dimensional Euclidian rotation – i.e., an element of $\mathfrak{so}(4)$ – and therefore $\alpha^{\alpha}{}_{\mu} \in \mathfrak{so}(3,1)$.

We now see that the Lie equation (6.5) takes the form of the conformal Killing equation for the Minkowski space. This allows us to infer the established inventory of known spacetime symmetries of the Maxwell system (cf., Bateman and Cunningham [**1, 2**]) since the vector fields on $\mathbb{R}^4$ that we get as its solutions are the infinitesimal generators of translations, homotheties, Lorentz transformations, and inversions through the light cone, as discussed in the previous section. The allowable transformations of the fiber of $\Lambda^2(\mathbb{R}^4)$ remain unchanged, however.

### 6.2 Non-uniform linear case

In this case:

$$d\kappa_{\mu\nu}{}^{\alpha\beta} = \kappa_{\mu\nu}{}^{\alpha\beta}{}_{,\lambda} dx^{\lambda}, \tag{6.19}$$

and the basic equations take the form:

$$0 = \Theta^1 \equiv dF_{\mu\nu} \wedge dx^{\mu} \wedge dx^{\nu}, \tag{6.20a}$$
$$0 = \Theta^2 \equiv \kappa_{\mu\nu}{}^{\alpha\beta} dF_{\alpha\beta} \wedge dx^{\mu} \wedge dx^{\nu} + F_{\alpha\beta} \kappa_{\mu\nu}{}^{\alpha\beta}{}_{,\lambda} dx^{\lambda} \wedge dx^{\mu} \wedge dx^{\nu}. \tag{6.20b}$$

Hence, the module $\mathcal{M}$ that is generated by $\Theta^1$ and $\Theta^2$ consists of 3-forms of the form:

$$(\alpha \, \delta^{\mu}{}_{\alpha} \delta^{\nu}{}_{\beta} + \beta \, \kappa_{\mu\nu}{}^{\alpha\beta}) \, dF_{\alpha\beta} \wedge dx^{\mu} \wedge dx^{\nu} + \beta \, F_{\alpha\beta} \, \kappa_{\mu\nu}{}^{\alpha\beta}{}_{,\lambda} \, dx^{\lambda} \wedge dx^{\mu} \wedge dx^{\nu}. \tag{6.21}$$

This differs from the previous case by the extra term involving the $dx^{\lambda} \wedge dx^{\mu} \wedge dx^{\nu}$. Hence, $X^{\mu} = X^{\mu}(x^{\nu})$, as always, but $X_{\mu\nu} = X_{\mu\nu}(x^{\alpha}, F_{\alpha\beta})$ now.

The infinitesimal symmetry equations for this case are obtained from (5.49a-g). The $X^{\mu}$ components must satisfy:

$$2\frac{\partial X^\alpha}{\partial x^\mu}\delta^\beta_{\ \nu} = \alpha_1 \delta^\alpha_{\ \mu}\delta^\beta_{\ \nu} + \beta_1 \kappa_{\mu\nu}{}^{\alpha\beta}, \tag{6.22a}$$

$$L_X \kappa_{\mu\nu}{}^{\alpha\beta} = \alpha_2 \delta^\alpha_{\ \mu}\delta^\beta_{\ \nu} + \beta_2 \kappa_{\mu\nu}{}^{\alpha\beta}, \tag{6.22b}$$

$$L_X \frac{\partial \kappa_{\mu\nu}{}^{\alpha\beta}}{\partial x^\lambda} = \gamma_2 \frac{\partial \kappa_{\mu\nu}{}^{\alpha\beta}}{\partial x^\lambda}, \tag{6.22c}$$

$$F_{\rho\sigma}(X^\lambda{}_{,\alpha}\,\kappa_{\beta\gamma}{}^{\rho\sigma}{}_{,\lambda} + X^\mu{}_{,\beta}\,\kappa_{\mu\gamma}{}^{\rho\sigma}{}_{,\alpha} + X^\nu{}_{,\gamma}\,\kappa_{\beta\nu}{}^{\rho\sigma}{}_{,\alpha}) = \gamma_3 F_{\rho\sigma}\,\kappa_{\alpha\beta}{}^{\rho\sigma}{}_{,\gamma}, \tag{6.22d}$$

and the $X_{\mu\nu}$ components must satisfy:

$$\frac{\partial X_{\mu\nu}}{\partial F_{\alpha\beta}} = \alpha_3\,\delta^\alpha_{\ \mu}\delta^\beta_{\ \nu} + \beta_3 \kappa_{\mu\nu}{}^{\alpha\beta}, \tag{6.22e}$$

$$\frac{\partial X_{\mu\nu}}{\partial x^\lambda} = \alpha\,F_{\alpha\beta}\frac{\partial \kappa_{\mu\nu}{}^{\alpha\beta}}{\partial x^\lambda}, \tag{6.22f}$$

$$X_{\alpha\beta}\frac{\partial \kappa_{\mu\nu}{}^{\alpha\beta}}{\partial x^\lambda} = \beta\,F_{\alpha\beta}\frac{\partial \kappa_{\mu\nu}{}^{\alpha\beta}}{\partial x^\lambda}. \tag{6.22g}$$

There is little that can be said in general about the nature of the $X^\mu$ components, since the $\kappa_{\mu\nu}{}^{\alpha\beta}$ are not constants, this time. One would generally have to look at further specializations in the nature of the electromagnetic constitutive law in order to find specific solutions.

As for the $X_{\mu\nu}$ components, since $\kappa$ is not a function of the $F_{\mu\nu}$, if we choose $\alpha_2$ and $\beta_2$ that are not functions of the $F_{\mu\nu}$, we can solve (6.22e) in the previous manner:

$$X_{\mu\nu} = \phi_{\mu\nu} + \alpha_3 F_{\mu\nu} + \beta_3 \kappa_{\mu\nu}{}^{\alpha\beta} F_{\alpha\beta}. \tag{6.23}$$

However, $X_{\mu\nu}$ must satisfy the constraints imposed by (6.22f, g). From (6.22f), we must have:

$$X_{\mu\nu,\lambda} = \phi_{\mu\nu,\lambda} + \alpha_{3,\lambda} F_{\mu\nu} + \beta_{3,\lambda}\kappa_{\mu\nu}{}^{\alpha\beta}F_{\alpha\beta} + \beta_3 \kappa_{\mu\nu}{}^{\alpha\beta}{}_{,\lambda}F_{\alpha\beta} = \alpha\,\kappa_{\mu\nu}{}^{\alpha\beta}{}_{,\lambda}F_{\alpha\beta} \tag{6.24}$$

for some $\alpha$. This immediately implies that:

$$\frac{\partial \phi_{\mu\nu}}{\partial x^\lambda} = \gamma\frac{\partial(\kappa_{\mu\nu}{}^{\alpha\beta}F_{\alpha\beta})}{\partial x^\lambda}F_{\alpha\beta}, \tag{6.25}$$

or, as long as $\gamma$ is not a function of $x^\lambda$:

$$\phi_{\mu\nu} = \gamma\,\kappa_{\mu\nu}{}^{\alpha\beta}F_{\alpha\beta}. \tag{6.26}$$

This reduces our symmetries to those of the form:

$$X_{\mu\nu} = \alpha_3 F_{\mu\nu} + \beta_3 \kappa_{\mu\nu}{}^{\alpha\beta}F_{\alpha\beta}. \tag{6.27}$$

Hence, the constraint (6.22f) has broken the symmetry that represents the addition of an arbitrary solution, although the system is still linear.

A solution to (6.24) for which $\alpha_3$ and $\beta_3$ are not constant is possible if $\kappa_{\mu\nu}{}^{\alpha\beta}$ satisfies:

$$\kappa_{\mu\nu}{}^{\alpha\beta}{}_{,\lambda} = \alpha_\lambda \, \delta^\alpha_\mu \, \delta^\beta_\nu + \beta_\lambda \, \kappa_{\mu\nu}{}^{\alpha\beta}, \tag{6.28}$$

for some $\alpha_\lambda$, $\beta_\lambda$, which must then be functions of only $x^\lambda$.

As for (6.22g), it entails that:

$$(\alpha_3 \, F_{\alpha\beta} + \beta_3 \kappa_{\alpha\beta}{}^{\kappa\lambda} F_{\kappa\lambda}) \kappa_{\mu\nu}{}^{\alpha\beta}{}_{,\lambda} = \beta \, F_{\alpha\beta} \, \kappa_{\mu\nu}{}^{\alpha\beta}{}_{,\lambda}, \tag{6.29}$$

which implies that:

$$\beta_3 = 0. \tag{6.30}$$

Hence, the remaining form for the $X_{\mu\nu}$ components of $X$ is:

$$X_{\mu\nu} = \alpha_3 F_{\mu\nu}, \tag{6.31}$$

with $\alpha_3$ constant, which leaves only the dilatation symmetry. Apparently, the constraint (6.22g) has broken the duality symmetry, which is plausible, since the definition of duality is traceable to the introduction of $\kappa_{\mu\nu}{}^{\alpha\beta}$, to begin with.

### 6.3 Uniform nonlinear case

In this case:

$$d\kappa_{\mu\nu}{}^{\alpha\beta} = \kappa_{\mu\nu}{}^{\alpha\beta,\lambda\tau} dF_{\lambda\tau}, \tag{6.32}$$

and the basic equations take the form:

$$0 = \Theta^1 = dF_{\mu\nu} \wedge dx^\mu \wedge dx^\nu, \tag{6.33a}$$
$$0 = \Theta^2 = K_{\mu\nu}{}^{\alpha\beta} \, dF_{\alpha\beta} \wedge dx^\mu \wedge dx^\nu. \tag{6.33b}$$

into which we have introduced the "deformed" electromagnetic constitutive law:

$$K_{\mu\nu}{}^{\alpha\beta} = \kappa_{\mu\nu}{}^{\alpha\beta} + F_{\kappa\lambda} \, \kappa_{\mu\nu}{}^{\kappa\lambda,\alpha\beta}, \tag{6.34}$$

which is then only a function of $F_{\mu\nu}$. Note that (6.33a, b) is still of Maxwellian form, only with a deformed constitutive law. Since it is conceivable that the equation:

$$K_{\mu\nu}{}^{\alpha\beta} = 0, \tag{6.35}$$

might have non-trivial solutions for $F_{\kappa\lambda}$ for some choices of $\kappa_{\mu\nu}{}^{\kappa\lambda}$, which would change the rank of the module $\mathcal{M}$ at such $F_{\kappa\lambda}$, it possible that the system (6.33a, b) might be singular at some points of $\Lambda^2(\mathbb{R}^4)$.

The module $\mathcal{M}$ generated by $\Theta^1$ and $\Theta^2$ now consists of terms of the form:

$$\alpha \Theta^1 + \beta \Theta^2 = [\alpha \, \delta^\alpha{}_\mu \, \delta^\beta{}_\nu + \beta \, K_{\mu\nu}{}^{\alpha\beta}] \, dF_{\alpha\beta} \wedge dx^\mu \wedge dx^\nu \,. \tag{6.36}$$

Again, this module itself may take on a different character at some $F_{\kappa\lambda}$ depending upon the nature of $\kappa_{\mu\nu}{}^{\alpha\beta}$.

The infinitesimal symmetry equations (5.51a-c) now take on the form:

$$2 \frac{\partial X^\alpha}{\partial x^\mu} \delta^\beta{}_\nu = \alpha_1 \delta^\alpha{}_\mu \delta^\beta{}_\nu + \beta_1 K_{\mu\nu}{}^{\alpha\beta}, \tag{6.37a}$$

for the $X^\lambda$ components of $X$, and:

$$\frac{\partial X_{\mu\nu}}{\partial F_{\alpha\beta}} = \alpha_3 \, \delta^\alpha{}_\mu \, \delta^\beta{}_\nu + \beta_3 K_{\mu\nu}{}^{\alpha\beta}, \tag{6.37b}$$

$$X_{\alpha\beta} \frac{\partial \kappa_{\mu\nu}{}^{\alpha\beta}}{\partial F_{\kappa\lambda}} = \alpha \delta^\lambda{}_\mu \, \delta^\kappa{}_\nu + \beta \, K_{\mu\nu}{}^{\kappa\lambda} \,. \tag{6.37c}$$

for the $X_{\mu\nu}$ components.

We observe that these equations are of the same form as we obtained for the case of constant $\kappa_{\mu\nu}{}^{\alpha\beta}$, except that the deformed tensor field $K_{\mu\nu}{}^{\alpha\beta}$ is not constant, but a function of $F_{\mu\nu}$. One must, however, note that the Lie derivative of the undeformed $\kappa$ now has to be coupled to the deformed tensor $K$.

Since $K_{\mu\nu}{}^{\alpha\beta}$ is not a function of $x^\mu$ we can integrate (6.37a) and obtain essentially the same solution that we derived for the case of uniform linear $\kappa^{\alpha\beta}{}_{\kappa\lambda}$, except that we are using $K^{\alpha\beta}{}_{\kappa\lambda}$ in its place now:

$$X^\mu = \varepsilon^\mu + \alpha x^\mu + K^\mu{}_\alpha x^\alpha + \text{(prolongation term)}. \tag{6.38}$$

In this expression, we are taking into account that there is a choice of four types of prolongations. Hence, we see that the effect of this type of nonlinearity in $\kappa^{\alpha\beta}{}_{\kappa\lambda}$ on the spacetime symmetries is by way of the deformation of the $\kappa^{\alpha\beta}{}_{\kappa\lambda}$.

The second equation (6.38b) has the appearance of a deformed version of the uniform linear symmetry equations, except that now $K^{\alpha\beta}{}_{\kappa\lambda}$ is assumed to be a function of $F_{\alpha\beta}$, so the integration is not as straightforward as before, and will generally depend upon the form of the electromagnetic constitutive law. We can, however, still set $\beta_2 = 0$ and solve:

$$\frac{\partial X_{\mu\nu}}{\partial F_{\alpha\beta}} = \alpha_2 \, \delta^\alpha{}_\mu \, \delta^\beta{}_\nu \tag{6.39}$$

to obtain:

$$X_{\mu\nu} = \varepsilon_{\mu\nu} + \alpha_2 F_{\mu\nu} \,. \tag{6.40}$$

Although this symmetry is suggestive of a linear system of partial differential equations, not the nonlinear one at hand, nevertheless, these symmetries must still be consistent with (6.37c), which may break these symmetries.

What remains to be evaluated is the equation:

$$\frac{\partial X_{\mu\nu}}{\partial F_{\alpha\beta}} = \beta_2 K_{\mu\nu}{}^{\alpha\beta}, \tag{6.41}$$

which governs the duality symmetry of the nonlinear pre-metric Maxwell system. Again, its solution must also be consistent with (6.37c). However, in the absence of more details concerning the nature of $\kappa_{\mu\nu}{}^{\alpha\beta}$, there is little that can be added at this point.

## 7 Discussion

Let us now summarize the results of the foregoing analysis from a more physical perspective:

The uniform linear case is a generalization of the one that corresponds to the classical electromagnetic vacuum, which is characterized by a uniform linear isotropic electromagnetic constitutive law. If we choose a rest frame then we can decompose the 2-form $F$ into a sum of an electric and magnetic term:

$$F = \theta \wedge E - {*}(\theta \wedge B), \tag{7.1}$$

in which $\theta$ is a timelike unit covector, $E$ and $B$ are spacelike 1-forms, and $*$ is the usual Hodge dual isomorphism that one derives from the Lorentzian structure on the spacetime manifold. The electromagnetic excitation that is associated with $F$ for this vacuum is then:

$$H = \varepsilon_0\, \theta \wedge E - \frac{1}{\mu_0} {*}(\theta \wedge B). \tag{7.2}$$

If one wishes to eliminate the dependency on the metric in this expression, one needs to do two things: account for the splitting of $\Lambda^2(M)$ into two complementary three-dimensional sub-bundles in manner that is independent of the notion of a timelike vector field, and eliminate the reference to the Hodge dual. Both objectives can be accomplished by passing from the geometry of $T(M)$ given a Lorentzian structure to the geometry of $\Lambda^2(M)$ given an almost-complex structure. This shift in emphasis also corresponds to a shift from Lorentzian geometry to complex projective geometry. In effect, it is analogous to shifting from the Riemannian viewpoint that geometry is something that derives from a metric to the Kleinian viewpoint that "projective geometry is all geometry."

There are plausible physical reasons for making this conceptual transition. Firstly, the spacetime metric seems to have essentially originated in the study of wavelike solutions to the electromagnetic field equations. Hence, it represents a sort of reduction in scope for the study of more general electromagnetic fields. Indeed, most of the physical discussions of how one goes about measuring lengths of things seem to crucially involve the use of light rays or waves. Furthermore, as pointed out above, generically, "most" 2-forms on a four-dimensional manifold have rank four, whereas wavelike 2-forms have rank two. Hence, in a sense, the non-wavelike fields, such as static fields, are pre-metric by nature.

This is why the fact that one of the four possible avenues of prolongation for the Lie algebra of spacetime symmetries in the case of uniform linear electromagnetic constitutive laws is to generalize from the conformal geometry of light cones to projective geometry seems encouraging.

It seems clear that in the "linear optical" case of a non-uniform linear electromagnetic constitutive law, one must introduce more specific cases in order to get more of a physical intuition for how the non-uniformity in the constitutive law affects the symmetries of the pre-metric Maxwell system. In particular, it seems to be the case in which the spacetime symmetries are most explicitly coupled to the fiber symmetries.

One can regard the uniform nonlinear case of electromagnetic constitutive laws as either the "nonlinear optical" case, or a possible source of effective models for vacuum polarization in quantum electrodynamics that might extend the scope of existing models, such as Born-Infeld electrodynamics. It is indeed encouraging that the apparent effect of the nonlinear on the symmetries seems to be conceptually reasonable, although computationally involved. One can see how even spacetime geometry might be influenced by the electric and magnetic fields even in the simplest case of (7.2) if one replaces the constants with functions of $F$ – say $\varepsilon = \varepsilon(E)$, $\mu = \mu(B)$. One can induce a Lorentzian metric of the form:

$$\eta = -\frac{1}{\mu(B)}(\theta^0)^2 + \varepsilon(E)(\theta^i)^2 \tag{7.3}$$

on some four-dimensional subspaces of $\Lambda^2(M)$ by restricting the scalar product on $\Lambda^2(M)$ that is given by $<F, G> = \mathcal{V}(F \wedge G)$. If one factors out $\varepsilon(E)$ in this expression then one obtains:

$$\eta = \varepsilon(E) \left[-\frac{1}{c^2}(\theta^0)^2 + (\theta^i)^2\right], \qquad c^2 = \frac{1}{\varepsilon(E)\mu(B)}, \tag{7.4}$$

and one sees how the conformal geometry of the linear wave equation can emerge from assumptions about the linear electromagnetic constitutive law, which, however, break down for nonlinear constitutive laws. In particular, it would no longer be advisable to set $c$ equal to one, since it would not be constant, except approximately for electromagnetic fields strengths that are less than some excessive value that probably only becomes significant in precisely those phenomena that signaled the shift from Maxwellian electrodynamics to quantum electrodynamics, such as the behavior of electrostatic fields in the small neighborhoods of elementary charge distributions and the nonlinear scattering of photons by nuclear fields and other photons.

Since there seems to be a natural limit imposed upon the deformation of $\kappa$ into $K$ by way of the contribution from the nonlinearity in $\kappa$, it seems natural to investigate how the breakdown of invertibility for the map that $K$ defines might be associated with a corresponding mathematical description of the aforementioned limiting regime for $F$.

**Acknowledgements** The author wishes to thank B. Kent Harrison and George Bluman for illuminating discussions on the symmetries of differential equations, as well as Anatoly Nikitin for organizing and hosting the Kiev Symmetry in Nonlinear Physics conference in the Summer of 2005, and to thank Bethany College for providing the opportunity to pursue theoretical research in an unhurried and intellectually supportive environment.